\documentclass[fleqn,usenatbib]{mnras}

\usepackage{newtxtext,newtxmath}

\usepackage[T1]{fontenc}
\usepackage{ae,aecompl}

\usepackage{graphicx}	
\usepackage{amsmath}	
\usepackage{amssymb}	
\usepackage{color}
\usepackage{syntonly}
\usepackage{float}

\title{The reverberation signatures of rotating disc winds in active galactic nuclei}

\author[S. W. Mangham et al.]{
S. W. Mangham,$^{1}$\thanks{E-mail: s.w.mangham@soton.ac.uk}
C. Knigge,$^{1}$
J. H. Matthews,$^{2}$
K. S. Long,$^{3,4}$
S. A. Sim,$^{5}$ \newauthor
and N. Higginbottom$^{1}$
\\
$^{1}$ Department of Physics and Astronomy, University of Southampton, Southampton, SO17 1BJ, UK\\
$^{2}$ University of Oxford, Astrophysics, Keble Road, Oxford, OX1 3RH, UK\\
$^{3}$ Space Telescope Science Institute, 3700 San Martin Drive, Baltimore, MD 21218, USA\\
$^{4}$ Eureka Scientific, Inc., 2452 Delmer St., Suite 100, Oakland, CA 94602-3017, USA\\
$^{5}$ School of Mathematics and Physics, Queen's University Belfast, University Road, Belfast, BT7 1NN, UK}

\date{Accepted 2017 July 20. Received 2017 June 22; in original form 2016 December 6}

\pubyear{2017}

\begin{document}
\label{firstpage}
\pagerange{\pageref{firstpage}--\pageref{lastpage}}
\maketitle

\begin{abstract}
The broad emission lines (BELs) in active galactic nuclei (AGN) respond to ionizing continuum variations. The time and velocity dependence of their response depends on the structure of the broad-line region: its geometry, kinematics and ionization state. Here, we predict the reverberation signatures of BELs formed in rotating accretion disc winds. We use a Monte Carlo radiative transfer and ionization code to predict velocity-delay maps for representative high- (C~{\sc iv}) and low-ionization (H$\alpha$) emission lines in both high- and moderate-luminosity AGN. Self-shielding, multiple scattering and the ionization structure of the outflows are all self-consistently taken into account, while small-scale structure in the outflow is modelled in the micro-clumping approximation. Our main findings are: (1)~The velocity-delay maps of smooth/micro-clumped outflows often contain significant \emph{negative} responses. (2)~The reverberation signatures of disc wind models tend to be rotation dominated and can even
resemble the classic ``red-leads-blue" \emph{inflow} signature. (3)~Traditional ``blue-leads-red" outflow signatures can usually only be observed in the long-delay limit. (4)~ Our models predict lag-luminosity relationships similar to those inferred from observations, but systematically underpredict the observed centroid delays. (5)~The ratio between ``virial product" and black hole mass predicted by our models depends on viewing angle. Our results imply that considerable care needs to be taken in interpreting data obtained by observational reverberation mapping campaigns. In particular, basic signatures such as ``red-leads-blue", ``blue-leads-red" and ``blue and red vary jointly" are not always reliable indicators of inflow, outflow or rotation. This may help to explain the perplexing diversity of such signatures seen in observational campaigns to date.
\end{abstract}




\section{Introduction}
\label{sec:background}
Reverberation mapping (RM) has become a powerful tool in the study of
active galactic nuclei (AGN; \citealt{Peterson1993}). Its primary application to date has
been in efforts to estimate central black hole (BH) masses. The
broad emission lines (BELs) in AGN respond to variations in the underlying continuum with a
characteristic time delay, $\tau$. This delay is due to the light travel time from the
central engine to the broad line region (BLR) and therefore depends on the
size of the BLR, $R_{BLR} \simeq c \tau$. If the dynamics of the BLR
are dominated by the gravitational potential of the BH, the width of
the Doppler-broadened emission lines must be
$\Delta v_{BLR} \simeq (G M_{BH} / R_{BLR})^{1/2}$. Thus the combination of
line width and lag immediately provides an estimate of the
BH mass, the so-called {\em virial product} $M_{BH} \simeq c
\tau_{BLR} v_{BLR}^2  / G$.

Lag-based BH mass estimates have been successfully calibrated
against both the $M_{BH}-\sigma$ \citep{Gebhardt2000,Ferrarese2001,Onken2004,Woo2010} and the
$M_{BH}-M_{bulge}$ relations \citep{Wandel2002,McLure2002}. However, even though the BLR sizes
obtained via RM seem to yield reliable BH masses, the nature of the BLR itself has remained
controversial.
This uncertainty regarding the geometry and kinematics of the BLR also limits
the accuracy of RM-based BH mass estimates \citep{Park2012, Shen2014, Yong2016}.

RM itself offers one of the most promising ways to
overcome this problem, since the physical properties, geometry and
kinematics of the BLR are encoded in the time- and velocity-resolved
responses of BELs to continuum
variations. Decoding this information places strong demands on
observational data sets \citep{Horne2004}, but recent campaigns have
begun to meet these requirements (e.g. LAMP [\citet{Bentz2011,Pancoast2012, Pancoast2014,Skielboe2015}],
SEAMBH [\citet{Du2014,Du2016}]; AGN Storm [\citet{DeRosa2015}]). Most previous velocity-resolved RM studies found signatures that were interpreted as evidence for inflow and/or rotation \citep{Ulrich1996, Grier2013, Bentz2008, Bentz2010, Gaskell1988, Koratkar1989}, although apparent outflow signatures have also been reported \citep{Denney2009,Du2016}.

On the theoretical side, considerable effort has been spent over the
years on modelling and predicting BLR reverberation signatures
\citep[e.g.][]{Blandford1982, Welsh1991, Wanders1995, Pancoast2011, Pancoast2012, Pancoast2014}. However, most modelling
efforts to date have treated the line formation process by
adopting parameterized emissivity profiles, rather than calculating the
actual ionization balance and radiative transfer within the
BLR self-consistently. Also, one of the most promising models for the
BLR -- rotating accretion disc winds \citep{Young2007} -- has received
surprisingly little attention in these modelling efforts.

Evidence for powerful outflows from luminous AGN is unambiguous:
approximately 10\% - 15\% of quasars display strong, blue-shifted
broad absorption  lines (BALs) in UV resonance transitions such as C~{\sc iv}~1550~\AA\ \citep[e.g.][]{Weymann1991,Tolea2002}. The {\em intrinsic} BAL fraction must be even higher
-- probably 20\% - 40\% -- since there are significant selection
biases against these so-called BALQSOs \citep{Hewett2003, Reichard2003, Knigge2008, Dai2008, Allen2011}. These outflows matter. First, they provide a natural
mechanism for ``feedback'', i.e. they allow a supermassive BH
to affect its host galaxy on scales far beyond its gravitational
sphere of influence \citep{Fabian2012}. Second, they can produce most of the
characteristic signatures of AGN, from BELs
\citep{Emmering1992, Murray1995} to BALs to X-ray warm absorbers \citep{Krolik1995}.
Third, they may be the key to unifying the diverse classes of AGN/QSOs \citep{Elvis2000}.

Despite the importance of these outflows, there have been few
attempts to study the reverberation signature of rotating disc
winds. \citet{Chiang1996} - hereafter CM96 - calculated the time-
and velocity-resolved response of the C~{\sc iv}~1550~\AA\ line based on
the line-driven disc wind model described by \citet{Murray1995}. Taking the outflow to be smooth, CM96 treated
velocity-dependent line transfer effects self-consistently, but
adopted a power-law radial emissivity/responsivity profile instead of carrying out
photoionization calculations. They also assumed that only the densest parts
of the wind (near the disc plane) produce significant line
emission. More recently, \citet{Waters2016} used a similar formalism
to predict the emission line response according to their hydrodynamic
numerical model of a line-driven AGN disc wind.

Similarly, \citet{Bottorff1997} predicted time-
and velocity-dependent response functions for C~{\sc iv}, based on the
centrifugally-driven hydromagnetic disc wind model of \citet{Emmering1992}.  In this model, the BLR is composed of
distinct, magnetically confined clouds, each of which is assumed
to be optically thick to both the Lyman continuum and C~{\sc iv} and
to possess no significant internal velocity shear. Moreover, \citet{Bottorff1997} adopted the
``locally optimally emitting cloud'' (LOC) picture of the BLR \citep{Baldwin1995},
according to which a population of clouds spanning a huge
range of particle density is assumed to exist at any given distance
from the central engine. These assumptions dramatically reduce the
complexity of the necessary photoionization and line transfer
calculations.

Against this background, our goal here is to use a fully
self-consistent treatment of ionization and radiative transfer to
predict H$\alpha$ and C~{\sc iv} reverberation signatures for a
generic model of rotating accretion disc winds in both high- and
moderate-luminosity AGN. Our starting point is the disc wind model
developed by \citet{Matthews2016} for luminous QSOs. This model was
explicitly designed with disc-wind-based geometric unification in mind:
it predicts strong BALs for observer orientations that look into the outflow,
produces strong emission lines for other orientations and is broadly
consistent with the observed X-ray properties of both QSOs and BALQSOs. We use
a scaled-down version of this model to represent lower-luminosity AGN, such as
Seyfert galaxies. For both types of model AGN, we calculate the time- and
velocity-resolved emission line responses and compare the predicted reverberation
signatures to observations.

The remainder of this paper is organized as follows. In section \ref{sec:Methods}, we outline the theory behind RM and describe the code we have employed to model it. In section \ref{sec:Biconical}, we use the code to calculate model response functions for both Seyfert galaxies and QSOs. In section \ref{sec:Discuss}, we test the lag-luminosity relationships predicted by our models against observations. We also present the viewing angle dependence of the virial product for these models and compare it to the results of recent observational modelling efforts. Finally, in section \ref{sec:Summary}, we summarize our conclusions.



\section{Methods}
\label{sec:Methods}

\subsection{Fundamentals}

\begin{figure}
	\centering
	\includegraphics[width=.75\linewidth, trim=1cm 2cm 1cm 1cm]{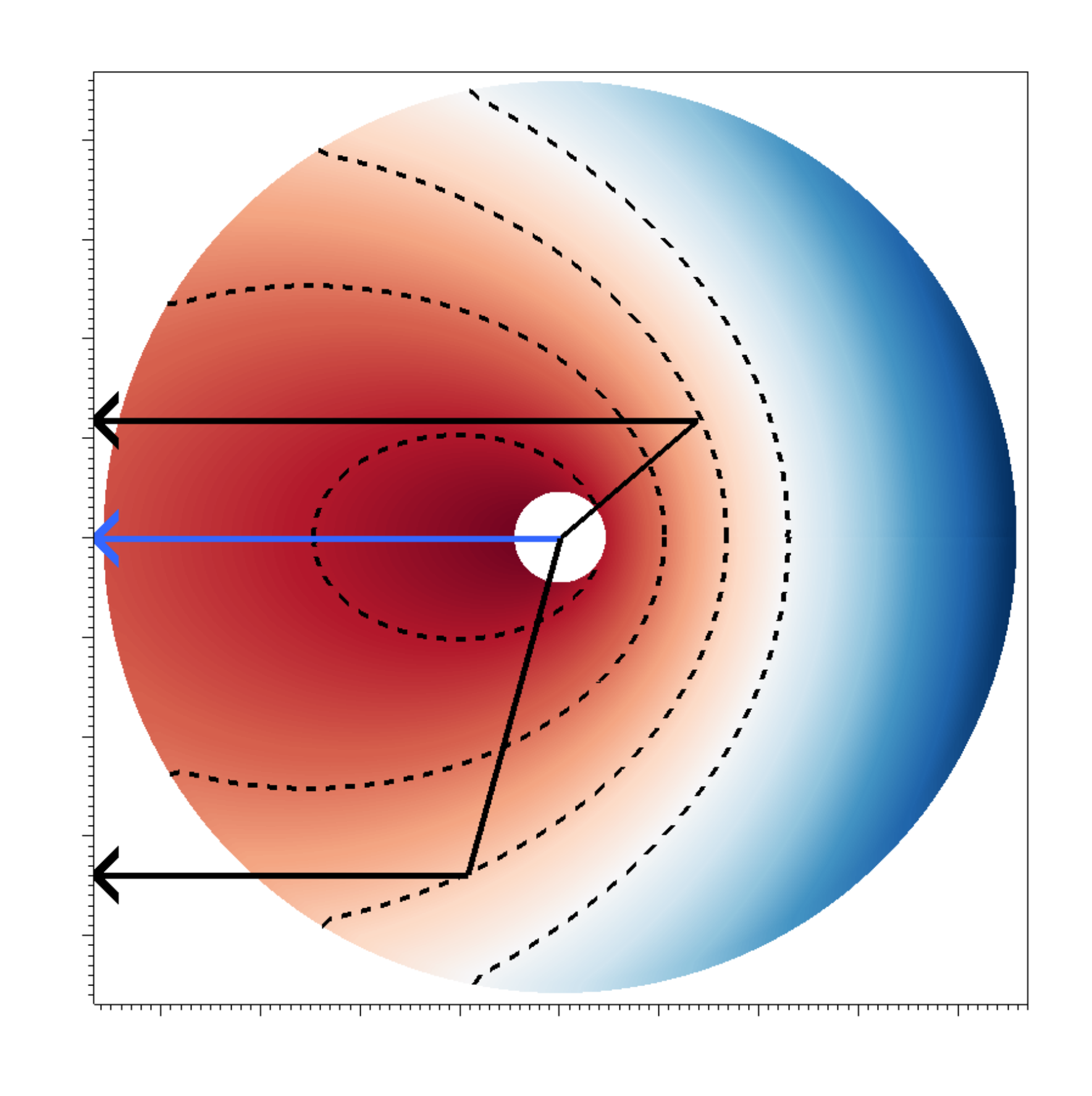}
	\caption{A cut through the isodelay surfaces surrounding a point source. Colour indicates total path to observer for photons scattering off that point; contours indicate isodelay curves on the 2-d plot. The blue arrow shows the direct path to observer; the two black arrows show the paths taken by two photons that scatter on different points of the same isodelay surface. Figure produced using Visit (\protect\cite{Visit}).}
	\label{fig:isodelay}
\end{figure}

The premise of RM is that the BLR reprocesses the
ionizing continuum generated by the central engine. This immediately implies
that the emission lines produced in the BLR should respond to variations
in the continuum. The response of a small parcel of BLR gas
to a change in the continuum flux it receives is effectively
instantaneous. As seen by a distant
observer, parts of the BLR that lie directly along the line of
sight to the central engine will therefore respond to a continuum
pulse with no delay. However, the response from any other
part of the BLR will be delayed with respect to the continuum. For
each location in the BLR, the delay is just the extra time it takes
photons to travel from the central engine to the observer, via this
location. As illustrated in Figure~\ref{fig:isodelay}, the isodelay
surfaces relative to a point-like continuum source are paraboloids
centered on the line of sight from the source towards the observer.

The response of a line will usually involve a
range of delays. The relationship between continuum variations,
$\Delta C(t)$, and variations in the flux of an emission line, $\Delta L(t)$,
can be expressed in terms of a {\em response function}, $\Psi_R(\tau)$,
\begin{equation}
	\Delta L(t) = \int_{-\infty}^{\infty} \Delta C(t - \tau) \Psi_R(\tau) d\tau.
	\label{eqn:Transfunc3}
\end{equation}
The response function describes how the emission line flux responds to a sharp
continuum pulse as a function of the time delay, $\tau$.

Since each parcel of gas has a different line of sight velocity $v$ with respect to the observer the line profile will also change with time.  
line of sight.
We can therefore define a 2-dimensional response
function, $\Psi_R(v,\tau)$, that specifies the response of the BLR as
a function of both time delay and radial velocity,
\begin{equation}
	\Delta L(v,t) = \int_{-\infty}^{\infty} \Delta C(t - \tau) \Psi_R(v,\tau) d\tau.
	\label{eqn:Transfunc2}
\end{equation}
While the 1-dimensional response function depends only on the
geometry of the BLR, the 2-dimensional response function -- also known as the ``velocity-delay map'' -- encodes information about both geometry and kinematics.

\begin{figure}
	\includegraphics[width=\columnwidth, trim=0cm 2cm 0cm 1cm]{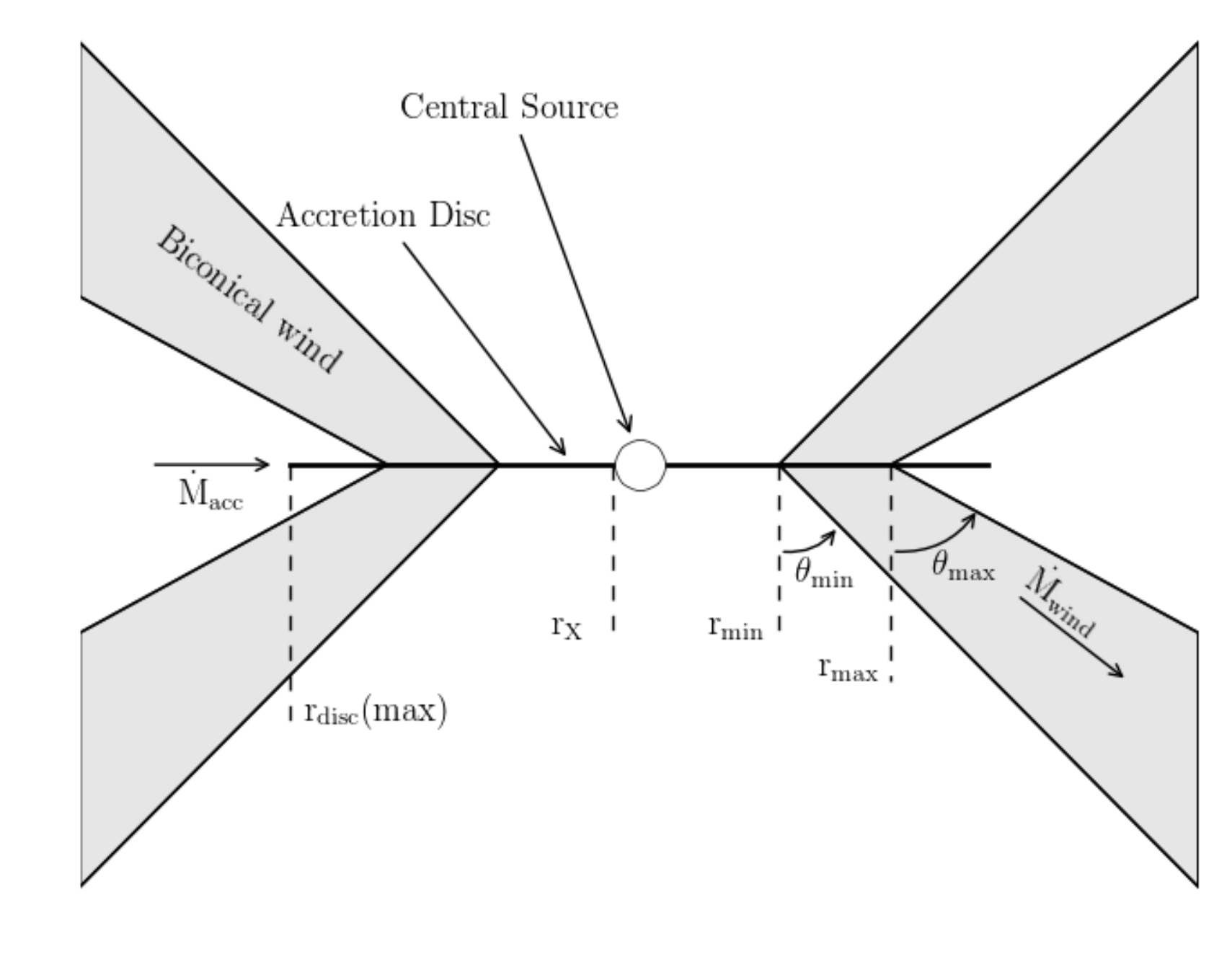}
    \caption{Sketch of the biconical disc wind geometry from \protect\cite{Matthews2016}.}
    \label{fig:model_james}
\end{figure}

Turning this argument around, we can {\em predict} the response of an
emission line to continuum variations for any physical model of the BLR.
We can test such models by comparing the response functions they
predict to those inferred from observations. Our goal here is to predict these
observational reverberation signatures for a rotating disc wind model of the BLR.
A sketch of the outflow geometry we adopt is shown in Figure~\ref{fig:model_james}, and is described in more detail in section \ref{sec:Biconical}.

\begin{figure}
	\includegraphics[width=\columnwidth, trim=0cm 1cm 0cm 1cm]{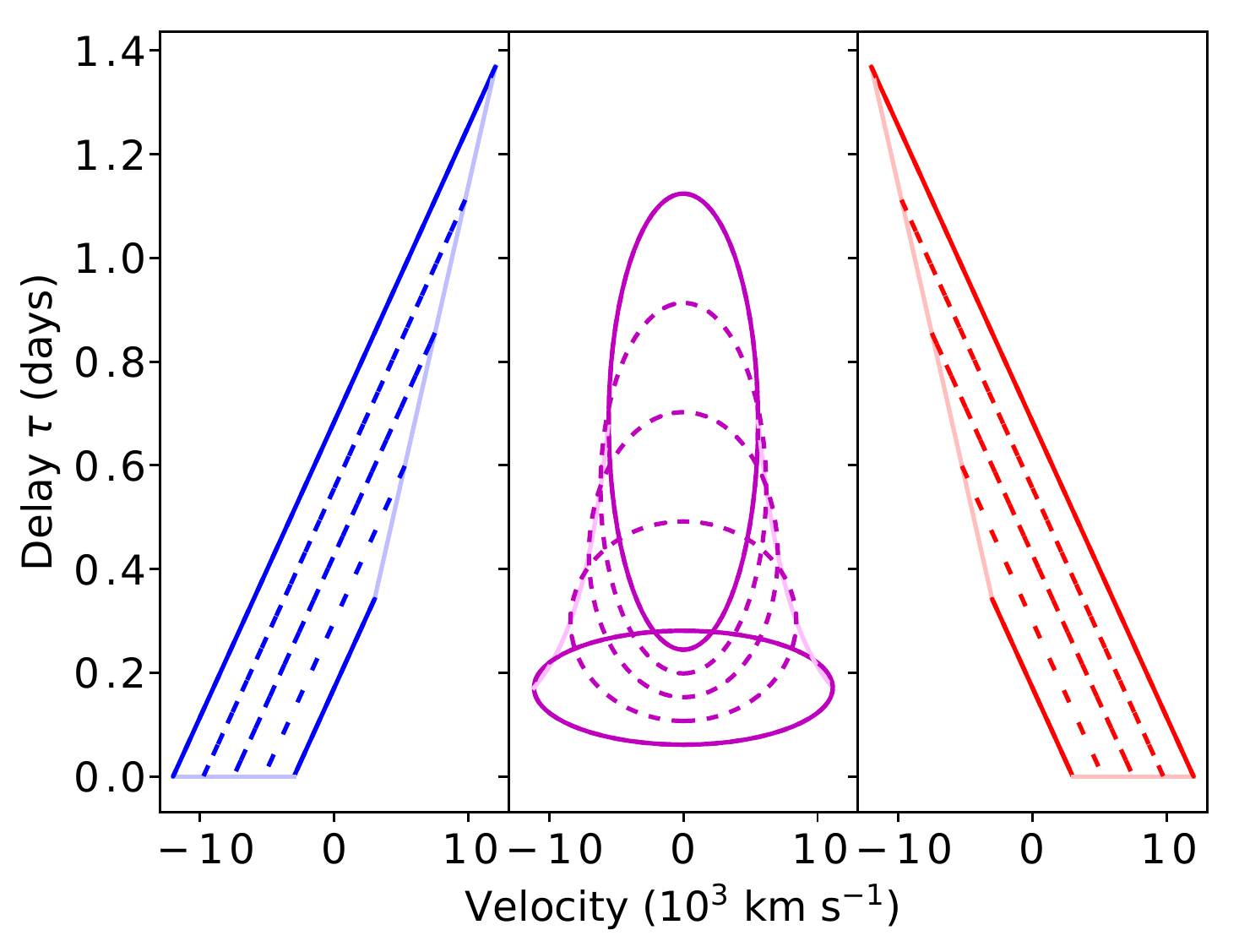}
    \caption{Outline response functions for Hubble-type spherical outflow \textcolor{blue}{\textbf{(left)}}, a rotating Keplerian disc viewed at a $20^{\circ}$ angle \textcolor{magenta}{\textbf{(centre)}}, and Hubble-type spherical inflow \textcolor{red}{\textbf{(right)}}. Winds extend from $r_{min}=20r_{g}$ to $r_{max}=200r_{g}$ for an AGN of mass $10^{7} M_{\odot}$. Hubble out/inflows have $V(r_{min})=\pm 3\times 10^{3}$ km s$^{-1}$. Solid lines denote the response from the inner and outer edges of the winds, dotted lines from evenly-spaced shells within the wind. Pale lines describe the edge of the velocity-delay shape of the response function.}
    \label{fig:tf_example}
\end{figure}

Our model includes both Keplerian rotation and outflow and may therefore be expected to show a mix of both rotation and outflow signatures. The conversion of spherical inflow, spherical outflow and rotating disc kinematics into velocity-delay space is illustrated in Figure~\ref{fig:tf_example} (for more details, see, e.\ g. \citet{Welsh1991}). Whilst a Keplerian disc has a symmetric signature, with the line wings leading the low-velocity core, outflow and inflow signatures are strongly asymmetric, with the blue wing leading the red for outflows, and the red wing leading the blue for inflows. As noted in section \ref{sec:background}, all of these basic signatures have been seen observationally, although symmetric and red-leads-blue signatures are more common.

Sketches like those shown in Figure~\ref{fig:tf_example} represent a purely geometric projection of BLR gas from 6-D position and velocity space onto our 2-D line-of-sight velocity and time delay space. The actual appearance of a velocity-delay map within the boundaries imposed by this projection depends on the {\em strength} of the line response across the map. As we shall see, this is a crucial effect that can make it difficult to identify even basic reverberation signatures with particular kinematics.

\subsection{Predicting reverberation signatures with self-consistent
  ionization and radiative transfer}

The ionization state of the BLR depends primarily on its density
structure, as well as on the luminosity and spectral shape of the
ionizing continuum. In an outflowing biconical disc wind, the
ionization structure is naturally stratified, which is in line with
observations \citep{Onken2004, Kollatschny2013, Kollatschny2014}.

For a given BLR ionization state, the emission
line formation process involves the generation of line photons
within the BLR and their radiative transfer through it. In order to
reliably predict the strengths, shapes and reverberation signatures of
BELs, it is necessary to deal correctly with radiative
transfer effects and solve self-consistently for the ionization state
of the BLR.

In our work, we accomplish this by using a modified version of our
Monte Carlo radiative transfer and ionization code \textsc{python} \citep{Long2002}, which
was developed specifically for modelling the spectra of accretion disc
winds. 
\textsc{python} is a useful tool for reverberation modelling,
since a given outflow geometry can be tested not just against the
response function of a single line, but also against the full
predicted spectrum. The code has already been described several
times in the literature \citep{Long2002, Sim2005, Noebauer2010,
Higginbottom2013, Higginbottom2014, Matthews2015, Matthews2016}, so we only
provide a brief overview here, focusing on the modifications we have made in
order to predict reverberation signatures. A full description of the
method, including tests against 
existing reverberation modelling, and an exploration of
radiative transfer effects in RM will be presented elsewhere.

\textsc{python} generates synthetic spectra for astrophysical systems
in which an outflow with given geometry, kinematics and density
structure reprocesses the radiation generated by one or more continuum
sources. Our AGN wind models are based on a kinematic
(i.e. parameterized) description of a rotating biconical outflow emanating
from the surface of the accretion disc \citep[Figure \ref{fig:model_james}]{Shlosman1993}.
The same description was used by \citet{Higginbottom2013} and \citet[hereafter M16]{Matthews2016}
to predict the spectroscopic signatures of disc winds in QSOs. The geometry and kinematics
of the wind are determined by a set of parameters (fully detailed in Table \ref{table:params}).
Our high-luminosity ("QSO") model is identical to that developed in M16, while our
lower-luminosity ("Seyfert") model is a simple scaled-down version of this. These models are described further in Section \ref{sec:Biconical}. All of our
radiative transfer and ionization  calculations are performed on a cylindrically symmetric
grid imposed on the wind. The relevant continuum sources are taken to be a geometrically thin,
optically thick accretion disc (which dominates the UV and optical light) and a compact,
optically thin corona (which dominates the X-ray emission).

The calculation of synthetic spectra with \textsc{python} proceeds in two
steps. In the first step, the code generates 'bundles' of continuum photons
across the full frequency range,
tracks their interactions and iterates towards thermal and
ionization equilibrium throughout the wind. In the second step, the
converged state of the wind is kept fixed, photons are generated only
across a limited frequency range, and emergent spectra are predicted
for a set of user-defined observer orientations.

In our models, reverberation delays are associated solely with photon light travel times. This is in line with other RM studies \citep{Welsh1991,Waters2016} and can be justified on the basis that the recombination time scales at
typical BLR densities (minutes to hours) are much shorter than light-travel times across the BLR (days to weeks).

For the purpose of this study, we have modified \textsc{python} to track
the distances travelled by photons that contribute to the observed
line emission. This is relatively straightforward, since
\textsc{python} is a Monte Carlo code that already explicitly follows
photons as they make their way through the outflow.

The use of a full ionization and radiative transfer code offers two
significant advantages in predicting reverberation signatures. First,
the dependence of line emissivity on position within the BLR is
calculated self-consistently, based on the density, ionization and
thermal structure of the outflow. Second, radiative transfer effects
are self-consistently taken into account. For example, a continuum (e.g. disc)
photon that scatters many times before reaching a particular
location in the BLR will have travelled a significantly longer
distance than one that reaches the same location without undergoing
any scatter. BLR locations that ``see'' primarily
scattered/reprocessed continuum photons will therefore respond to
continuum variations with a longer delay than expected solely based on geometry. A benefit
of this approach is that we can isolate the contributions of singly- or multiply-scattered
photons to the overall reverberation signature.

\subsection{Initializing and updating photon path lengths}

In line with essentially all RM models to date, we assume that all of
the correlated line and continuum variability we observe has its
origin in the immediate vicinity of the central BH. The
time-dependent signal produced there is then reprocessed by other
components in the system, such as the accretion disc and the BLR.

For the continuum photons produced by the accretion disc, we implement
this picture by assigning to each an initial path length equal to its
starting distance from the origin of the model, i.e. $r_{\mathrm{disc}} (\mathrm{min})$ from Table \ref{table:params} for photons emitted at the inner edge of the disc. For our QSO model, the UV emission of the accretion disc is concentrated within a radius several times smaller than the wind launch radius, whilst optical emission is concentrated within the radius of the wind base. Both of these lie within the radius within which the accretion disc temperature profile would be expected to be dominated by the central source luminosity \citep{King1997} and therefore would lag behind changes in the central source continuum. Photons produced by the X-ray emitting corona are assigned initial path lengths equal to $r_X = r_{\mathrm{disc}}(\mathrm{min})$, i.e. we assume that the corona varies coherently. As the corona is compact in comparison to the wind (at least a factor of 50 smaller), this is a reasonable approximation. The light travel time across the corona $2Rr_X/c$ is $\simeq 10$~min for the Seyfert model and $\simeq 16$~hrs for the QSO model.

\textsc{python} enforces thermal equilibrium throughout the outflow,
i.e. the heating and cooling rates are always in balance (in a converged model). Except for cooling
due to adiabatic expansion, the wind is also assumed to be in radiative equilibrium, i.e.
photon absorption and emission provide the only channels for net energy flows into and out of
a given grid cell. The effective initial path lengths of photons representing thermal wind emission
must therefore reflect the path length distribution of the absorbed
photons that were responsible for heating the wind. In order to ensure this,
\textsc{python} keeps track of the path length distribution of photons depositing energy into each cell. When a  `bundle' of photons with a given wavelength passes through a cell, its total energy is reduced in proportion to the optical depth it encounters along its path. We record its path, and the bundle's total energy, in the cell. When a
line photon is thermally emitted within a given grid cell, it is then
assigned an initial path length drawn from the distribution of paths contributing to heating weighted by the energy contribution at each path. We
refer to this as the thermal approach to path length initialization.

Most of the strong metal lines seen in AGN -- e.g. N~{\sc
v}~1240~\AA, Si~{\sc IV}~1440~\AA, C~{\sc iv} 1550~\AA -- are
collisionally excited. In \textsc{python}, the formation of these
lines is treated via a simple two-level atom approach \citep{Long2002,Higginbottom2013}.
This works particularly well for
resonance lines (such as all those listed above), in which the lower
level is the ground state of the ion. Since collisional excitation is
a thermal process, we use the thermal approach to path length initialization for all photons emitted in this way. Observations of these lines will also include a smaller contribution from continuum photons scattered into the line. These photons' path lengths are calculated purely on the basis of their origin and their travel through the winds.

The two-level atom approximation is not appropriate for lines in which the upper level either does not couple strongly to the ground state, or is primarily populated from above.
The most important example of such lines are the Hydrogen and Helium
recombination lines that are prominent in the UV and optical
spectra of AGN. To model such features more
accurately, one or both of these elements can be treated via Lucy's
\citep{Lucy2002,Lucy2003} {\em macro-atom} approach within \textsc{python} \citep{Sim2005,Matthews2015,Matthews2016}.
In this approach, a H or He
macro-atom can be ``activated'' by a photon packet travelling through a wind
cell, then undergoes a number of internal energy level ``jumps'', and
finally ``de-activates'' via the emission of a line or continuum photon.
In order to predict the
reverberation signature of a given H or He recombination line, the
code keeps track of all macro-atom de-activations associated with this line in a
given cell. After each de-activation, the path length travelled by the photon
that activated the macro-atom is stored, gradually building up a distribution.
When the same cell emits a photon associated with this recombination line, the
initial path length is then drawn from this distribution. This approach is
appropriate if photoionizations are quickly followed by radiative
recombinations in the BLR. We therefore refer to this as the
prompt recombination approach to path length initialization.

Our simulations allow for clumpiness in the wind within the ``micro-clumping" approximation \citep[e.g.][]{Hillier91,Hamann98,Matthews2016}.
Thus we assume that the outflow, though smooth
on a bulk scale, comprises very many small, optically thin clumps. These clumps enhance physical processes with a density-squared dependence. However, there is no effect on the photon paths due to micro-clumping compared to a fully smooth wind flow. Since each clump is optically thin at all wavelengths in this approximation, it is characterized by a unique set of physical conditions (ionization state, temperature, electron density...). Thus, like fully smooth flows, micro-clumped flows are stratified only on large scales. This is a fundamental difference to BLR models based on optically thick clouds, such as ``locally optimally-emitting cloud" (LOC) models \citep{Baldwin1995, Ferguson1997}.

\subsection{Generating Response Functions}

The response function, $\Psi_R(v, \tau)$, describes the time- and velocity-dependent response of an emission line to a continuum pulse. The contribution of a particular parcel of BLR gas to the overall response depends primarily on three factors: (i) its distance from the continuum source (which sets the characteristic delay with which it responds); (ii) its line-of-sight velocity (which sets the velocity at which we observe its response); (iii) its {\em responsivity} (which sets the strength of its response).

The responsivity is the hardest of these factors to calculate. It is a measure of the number of {\em extra} line photons that are created when a given {\em extra} number of continuum photons are produced. Note that this is different from the {\em emissivity} of the parcel, which describes the efficiency with which it  reprocesses continuum into line photons in a steady state. To see this more clearly, consider a parcel with line-of-sight velocity $v$ and characteristic delay $\tau$, which produces a line flux $L(v,\tau)$ when subjected to a steady continuum flux $C$. The reprocessing efficiency of this parcel can be defined simply as $\epsilon(v, \tau) = L(v, \tau) / C$. However, its responsivity measures how the line flux {\em changes} when the continuum does.

In the limit of small continuum variations, the overall response of the parcel is fully characterized by the partial derivative of the line flux it produces, $L(v, \tau)$, with respect to the continuum flux, $C$. If we adopt a power-law approximation for the dependence of $L(v,\tau)$ on $C$,
\begin{equation}
L(v, \tau) = L_0(v, \tau) \left(\frac{C}{C_0}\right)^\eta,
\end{equation}
we can evaluate this partial derivative and write the response function at $C = C_0$ in the form
\begin{equation}
\Psi_R(v, \tau) = \left.\frac{\partial L(v,\tau)}{\partial C}\right|_{C=C_0} = \eta \frac{L_0(v, \tau)}{C_0} = \eta \epsilon_0(v, \tau).
\end{equation}
The response function is therefore the product of reprocessing efficiency, $\epsilon_0(v,\tau)$, and a dimensionless responsivity, $\eta$.

In the limit that $\eta$ is constant throughout the BLR, the response function is just a scaled version of $\epsilon_0(v, \tau)$. We will refer to a response function calculated under the assumption that $\eta \equiv 1$ as an
{\em emissivity-weighted response function} ({\em EWRF}, originally defined as such in \citet{Goad1993}), $\Psi_E(v,\tau)$, i.e.
\begin{equation}
\Psi_E(v, \tau) = \epsilon(v, \tau) = \frac{L(v,\tau)}{C}.
\end{equation}
However, in general, the assumption that $\eta = {\rm constant}$ across the BLR is poor \citep[e.g.][] {Goad1993,OBrien1995,Korista2004,Goad2014}, i.e. $\eta = \eta(v, \tau)$. For example, a parcel of gas might be very efficient at reprocessing continuum into line photons, but may respond to a pulse of additional continuum photons by producing {\em fewer} line photons, either in relative or perhaps even in absolute terms. This is possible, for example, because its ionization state may have changed as a result of the pulse. Such a parcel will have a high emissivity (high reprocessing efficiency), but low (or even negative) responsivity.

Despite this, the emissivity-weighted response function provides a useful approximation for, and stepping stone towards, the response function proper. In our case, an EWRF can be calculated from a single run of {\sc python},  simply by keeping track of the delay associated with each photon arriving at the observer. For example, given a simulation for continuum luminosity $C_0$, we bin all photons from the simulation that are associated with a given line in time-delay- and velocity-space, yielding the 2-D EWRF for the line in that system.

In order to estimate the {\em true} (responsivity-weighted) response function, we then calculate two more EWRFs, for continuum luminosities $C_{\mathrm{low}} < C_0 < C_{\mathrm{high}}$, where $\Delta C = C_{\mathrm{high}} - C_{\mathrm{low}} \ll C_0$. In this limit, the line response is approximately linear, i.e. $\eta$ does not depend on $C$. This allows us to estimate the response function from the high-and low-state EWRFs as
\begin{align}
\Psi_R(v, \tau) &= \left.\frac{\partial L(v,\tau)}{\partial C}\right|_{C=C_0}
\simeq  \frac{\Delta L(v, \tau)}{\Delta C} \\
&\simeq \frac{L_{\mathrm{high}} - L_{\mathrm{low}}}{C_{\mathrm{high}} - C_{\mathrm{low}}}\\
&\simeq
\frac{  C_{\mathrm{high}} \Psi_{E, \mathrm{high}}(v, \tau) -
C_{\mathrm{low}}  \Psi_{E, \mathrm{low}}(v, \tau)  }
{ C_{\mathrm{high}} - C_{\mathrm{low}} }.
\label{eqn:Respfunc1}
\end{align}
For the response function calculations in this work, we use $\Delta C/C_0 = 0.1$. This is large enough to be astrophysically interesting, but small enough to ensure our models have an approximately linear response to changes in the continuum level.

\section{Results: the reverberation signatures of rotating accretion disc winds}
\label{sec:Biconical}

We are now ready to study the predicted reverberation signatures of our
simple disc wind scenario for the BLR. The basic geometry of our model
is shown in Figure~\ref{fig:model_james} and describes a
biconical outflow emerging from the surface of the accretion disc
around the supermassive black hole (SMBH). The outflow gradually accelerates away
from the disc plane, with each streamline reaching a terminal velocity
on the order of the local escape speed at the streamline's footpoint
on the disc surface. Similarly, material in the outflow initially
shares the (assumed Keplerian) rotational velocity in the accretion
disc, and then conserves its specific angular momentum as it rises
above the disc and moves away from the rotation axis. Much more
detailed descriptions of this basic disc wind model, including the importance of clumping within the wind, can be found in \citet{Shlosman1993}, \citet{Long2002}, \citet{Higginbottom2013},
 and \citet{Matthews2015,Matthews2016}.

To systematically study the reverberation signatures
predicted by this model for the BLR, we need to distinguish between at
least two types of AGN and two types of emission lines. First, the
highest luminosity AGN -- i.e. quasars/QSOs -- are known to have
significantly longer characteristic reverberation lags than lower
luminosity AGN -- i.e. Seyfert galaxies \citep{Grier2012, Kaspi2005}. Second,
low-ionization and high-ionization lines are also known to display
significantly different reverberation lags \citep[e.g.][]{Kaspi1999, Grier2012a}. Both of these
effects are qualitatively consistent with the idea that the physical
conditions within the BLR are set by photoionization physics \citep{Bentz2009}. As discussed in the previous
section, strictly speaking we should also distinguish between
collisionally excited and recombination lines in this context.

To cover this parameter space, we explicitly consider two distinct AGN models:
one representative of high-luminosity QSOs, the other representative of lower-luminosity
Seyferts. The physical characteristics we adopt for these models are listed
in Table \ref{table:params}. Moreover, for each model, we present results for
two types of lines: one high-ionization, collisionally excited metal line
(C~{\sc iv}~1550~\AA) and one low-ionization recombination line (H$\alpha$). H$\alpha$ was chosen due to the strong H$\alpha$ response in our model allowing for clearer response function plots than H$\beta$, though both lines are commonly used in RM studies.

\begin{table}
	\footnotesize
	\caption{Model parameters}
	\label{table:params}
	\begin{tabular}{l l | c c }
    	\hline \hline
		Parameter 				& Symbol, units					& Seyfert & QSO	\\
		\hline
		SMBH mass 				& $M_{\mathrm{BH}}$, \emph{$M_{\odot}$} 	& $10^{7}$	& $10^{9}$ \\
		Accretion rate			& $\dot{M}_{\mathrm{acc}}$, \emph{$M_{\odot}$ yr$^{-1}$}		&	0.02 & 5 \\
																														&	$\dot{m}_{\mathrm{acc}}$, \emph{$\dot{M}_{\mathrm{Edd}}$}			& $\approx 0.1$	& $\approx 0.2 $ \\
		X-ray power-law			&								&		 	&			\\
   		index					& $\alpha_{X}$					& $-0.9$	& $-0.9$ 	\\
		X-ray luminosity		& $L_{X}$, \emph{erg s$^{-1}$}	& $10^{43}$	& $10^{45}$ \\
        X-ray source radius		& $r_{X}$, \emph{$r_{g}$}		& 6 & 6	\\
        						& $r_{X}$, \emph{cm}			& $8.8 \times 10^{12}$		& $8.8 \times 10^{14}$	\\
		Accretion disc radii	& $r_{\mathrm{disc}}$(min)		& $r_{X}$ & $r_{X}$ \\
								& $r_{\mathrm{disc}}$(max), \emph{$r_{g}$}	& 3400 & 3400 \\
								& $r_{\mathrm{disc}}$(max), \emph{cm}		& $5 \times 10^{15}$			& $5 \times 10^{17}$		\\
		Wind outflow rate 		& $\dot{M}_{w}$, \emph{$M_{\odot}$ yr$^{-1}$}		& 0.02 & 5 \\
		Wind launch radii		& $r_{\mathrm{min}}$, \emph{$r_{g}$}	& 300 & 300 \\
								& $r_{\mathrm{max}}$, \emph{cm}		& $4.4 \times 10^{14}$ 		& $4.4 \times 10^{16}$	\\
								& $r_{\mathrm{min}}$, \emph{$r_{g}$} 	& 600 & 600 \\
								& $r_{\mathrm{max}}$, \emph{cm}		& $8.8 \times 10^{14}$ 		& $8.8 \times 10^{16}$	\\
		Wind launch angles		& $\theta_{min}$				& $70^{\circ}$ & $70^{\circ}$ \\
								& $\theta_{max}$				& $82^{\circ}$ & $82^{\circ}$ \\
		Terminal velocity		& $v_{\infty}(r_0)$	 			& $v_{esc}(r_0)$ 				& $v_{esc}(r_0)$				\\
		Acceleration length		& $R_{v}$, \emph{cm}			& $10^{16}$	& $10^{19}$ \\
		Acceleration index 		& $\alpha$						& 1.0 & 0.5 \\
		Filling factor 			& $f_V$							& 0.01 & 0.01 \\
		Viewing angle 			& $i$							& 20$^{\circ}$ & 20$^{\circ}$ \\
		Number of photons		& 							& $2\times 10^{9}$ & $3\times 10^{9}$ \\ \hline \hline

\end{tabular}
\end{table}

\subsection{A high-luminosity QSO}

The key parameters of our representative QSO disc wind model are
summarized in Table~\ref{table:params}. It is essentially identical to the
model described in \citet{Matthews2016}, which was in turn based on
the model described by \citet{Higginbottom2013}. The \citet{Matthews2016}
 model is an attempt to test disc-wind based
geometric unification for BALQSOs and ``normal'' (non-BAL) QSOs.
It produces BALs for sightlines that look directly into the wind cone
and BELs for sightlines that do not. It is also broadly consistent
with the observed X-ray properties of both QSOs and BALQSOs.

\begin{figure}
	\includegraphics[height=\columnwidth, angle=-90, trim=0cm 0cm 0cm 0cm]{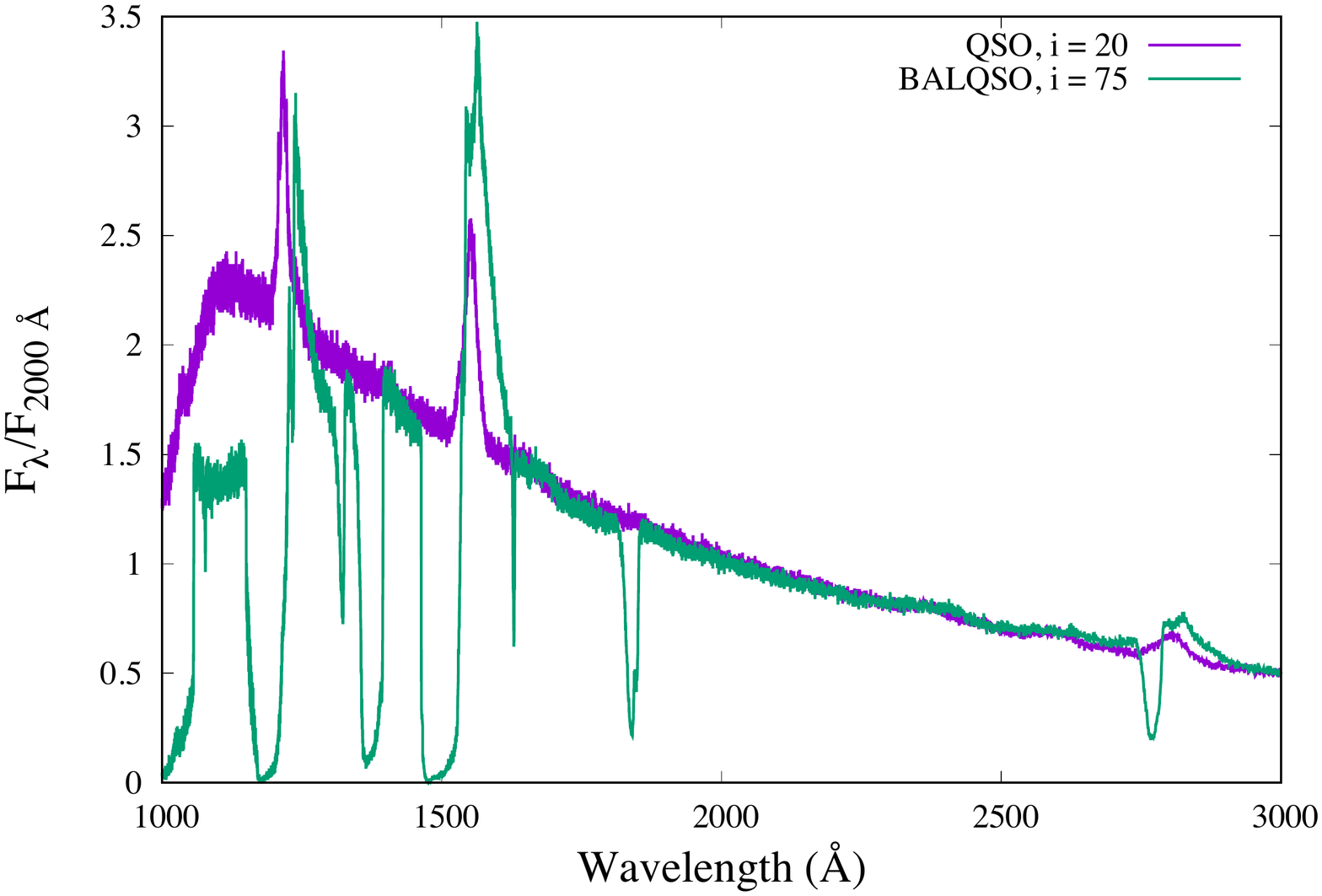}
    \caption{Spectra produced for the disc wind model of \protect\cite{Matthews2016}, viewed along QSO ($i = 20^{\circ}$) and BALQSO ($i=75^{\circ}$) sightlines. Both spectra have been normalised to unity at 2000~\AA.}
    \label{fig:spec_agn}
\end{figure}

Figure~\ref{fig:spec_agn} shows the UV and optical spectra predicted by this
model for two representative sightlines. The first corresponds to an
inclination of $i = 75^\circ$, where $i = 90$ corresponds to an
edge-on view of the system. An observer at this orientation
necessarily views the central engine through the accretion disc wind
and therefore sees a BALQSO. The second sight line corresponds to $i =20^\circ$
and allows a direct view of the inner accretion disc. An
observer at this orientation should therefore see the system as an
ordinary Type~I QSO. The spectra in Figure~\ref{fig:spec_agn} are consistent with
this. The predicted UV spectrum for $i = 75^\circ$ exhibits strong
blue-shifted BALs and P Cygni features, whereas the spectra for $i =
20^\circ$ only show BELs superposed on a blue continuum. Since absorption features
are necessarily formed along the line of sight to the continuum source, BAL troughs
respond to continuum variations without any light travel time delays. Our focus here is therefore
on the BEL reverberation signatures in our $i =20^\circ$ Type I QSO model.

\begin{figure*}
	\begin{minipage}{.48\textwidth}
		\includegraphics[width=\columnwidth, trim=0cm 0cm 0cm 0cm]{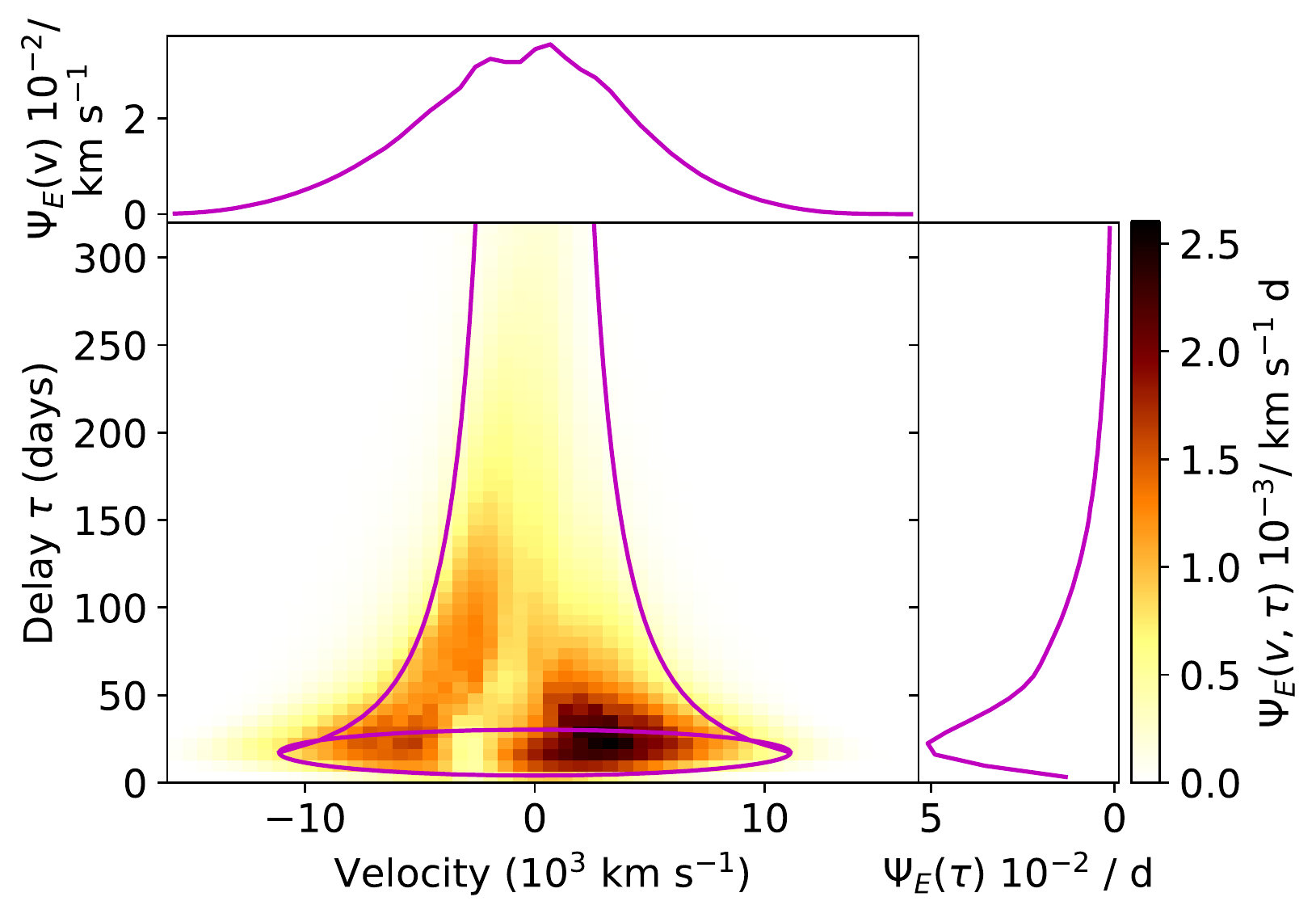}
    	\includegraphics[width=\columnwidth, trim=0cm 0cm 0cm 0cm]{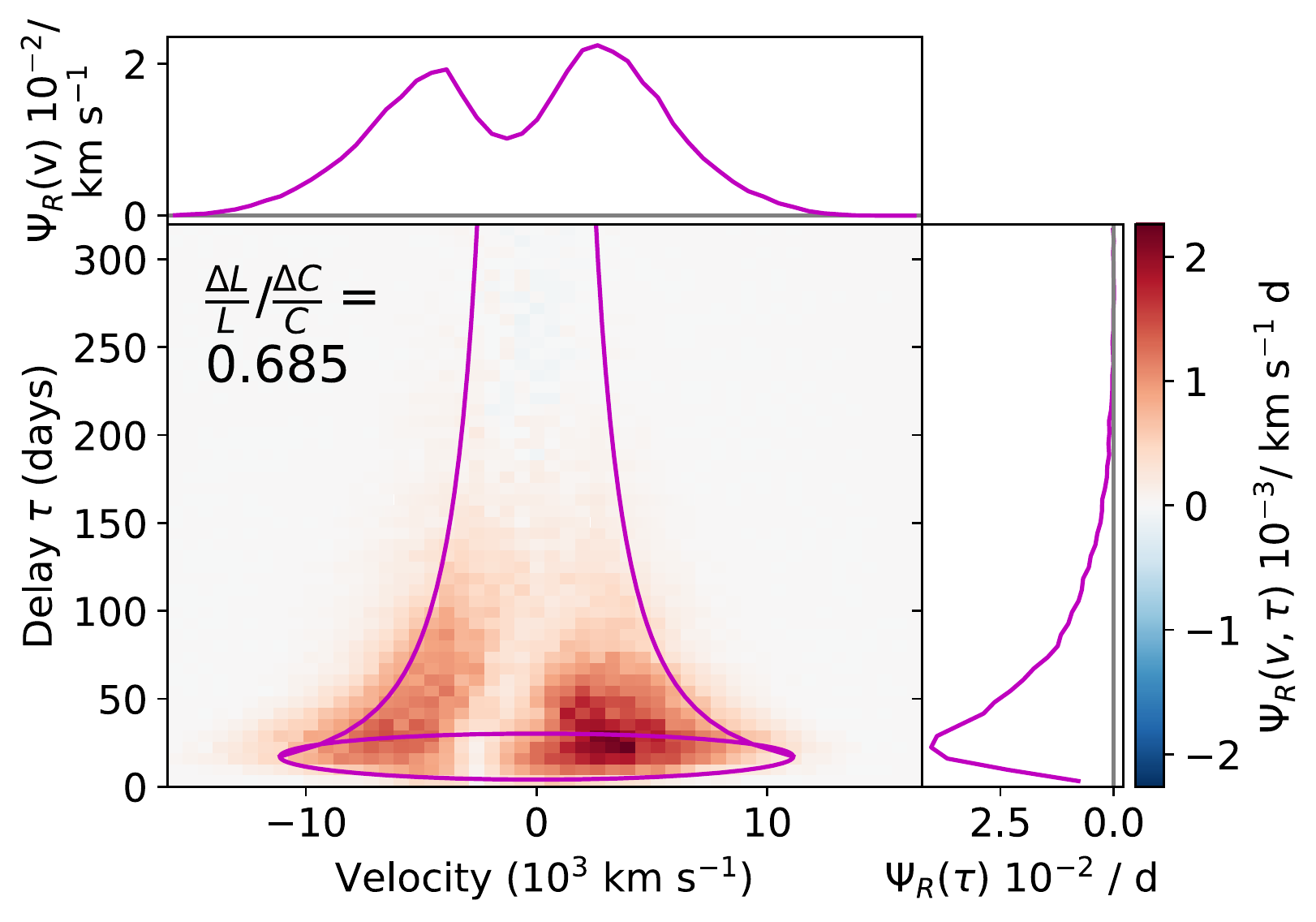}
    	\caption{The velocity-resolved EWRF (upper) and response function (lower) for H$\alpha$ in
      the QSO model, shown for time delays up to 320 days. The violet lines illustrate the function edges for a Keplerian disc at the wind launching radius, and the $\nu \propto \tau^{-1/2}$ "virial flow" envelope. Note that the velocity-integrated line response is weakly {\em negative} at low velocities beyond $\tau \gtrsim 220$~days.}
    	\label{fig:tf_agn_ha}
    \end{minipage}\hfill
    \begin{minipage}{.48\textwidth}
		\includegraphics[width=\columnwidth, trim=0cm 0cm 0cm 0cm]{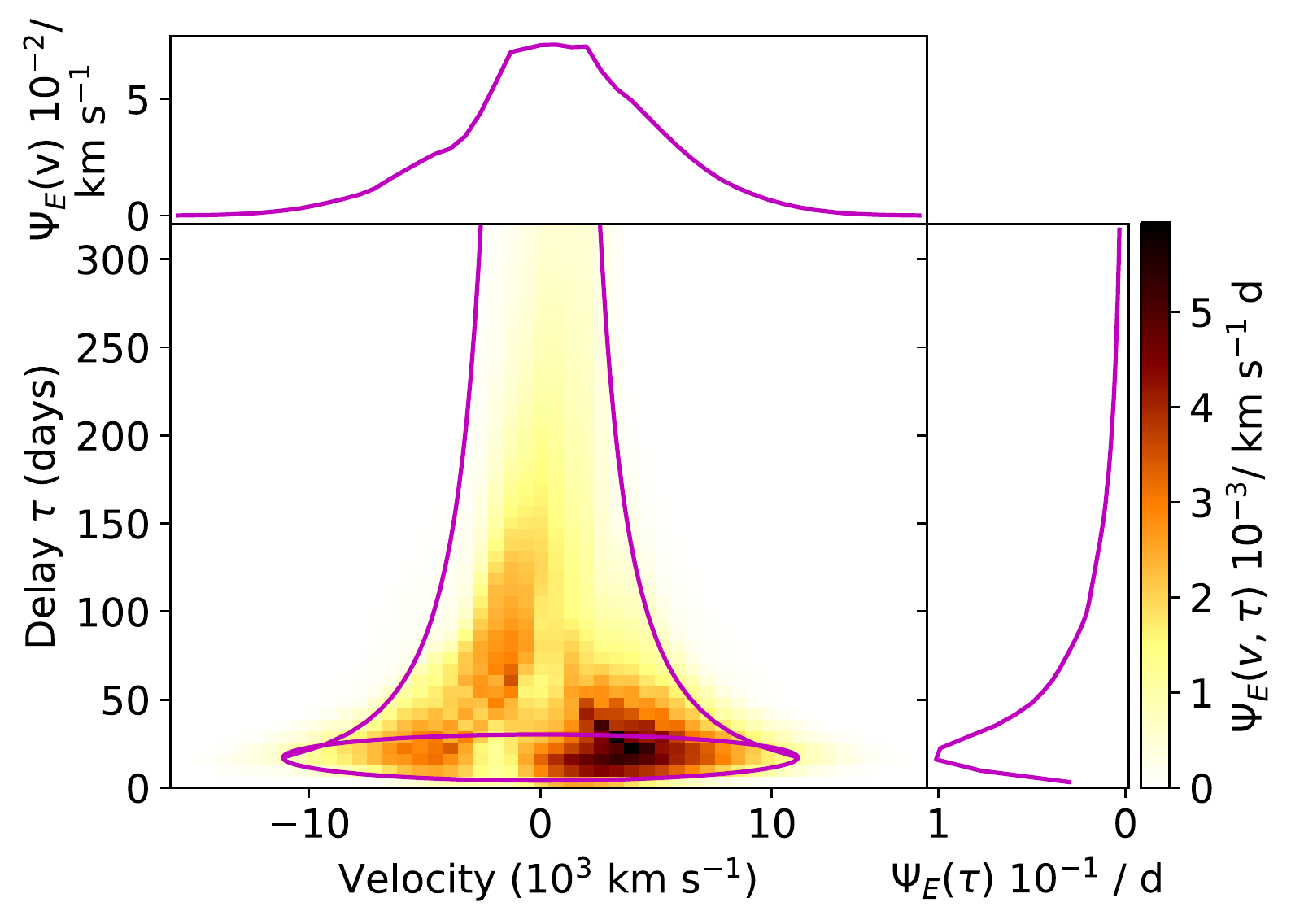}
    	\includegraphics[width=\columnwidth, trim=0cm 0cm 0cm 0cm]{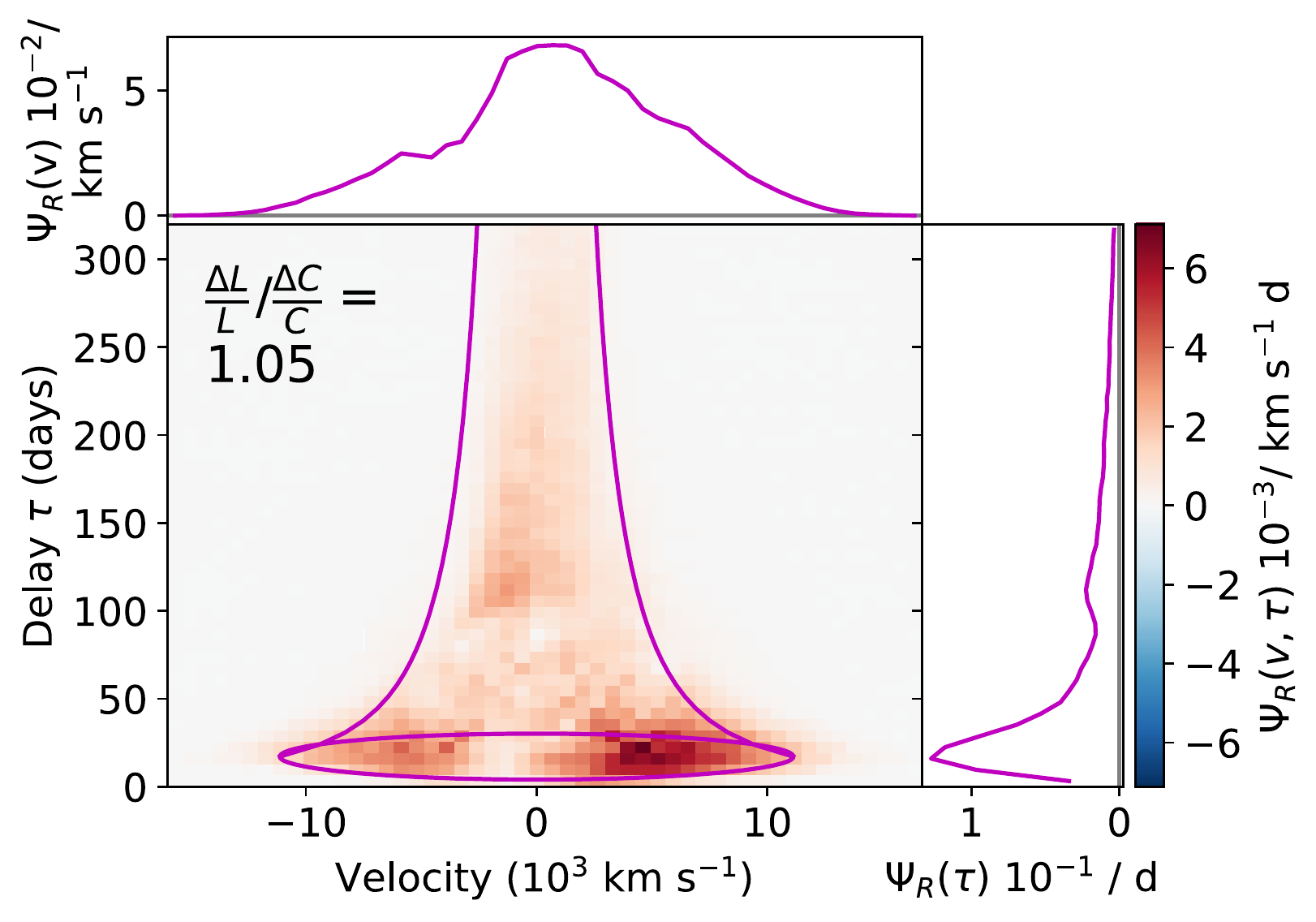}
   		\caption{The velocity-resolved EWRF (upper) and response function (lower) for C~\textsc{iv} in
      the QSO model, shown for time delays up to 320 days. The violet lines illustrate the function edges for a Keplerian disc at the wind launching radius, and the $\nu \propto \tau^{-1/2}$ "virial flow" envelope.}
    	\label{fig:tf_agn_c4}
    \end{minipage}

	\begin{minipage}{.48\textwidth}
      \includegraphics[width=\columnwidth, trim=2cm 4cm 2cm 2cm]{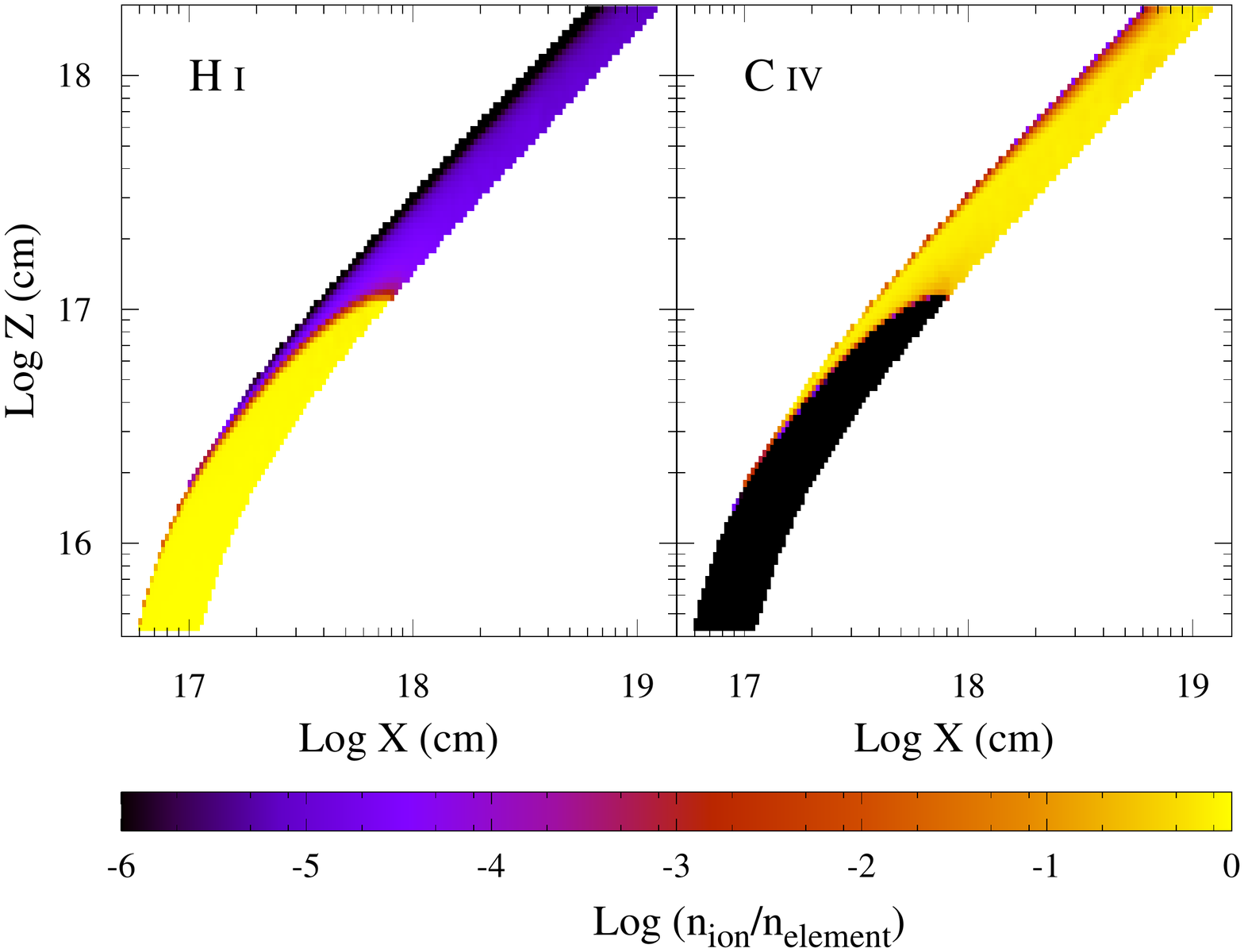}
      \caption{The H~\textsc{i} and C~\textsc{iv} ionization fractions within the wind in the QSO model.}
      \label{fig:agn_ions}
    \end{minipage}\hfill
    \begin{minipage}{.48\textwidth}
      \includegraphics[width=\columnwidth, trim=2cm 4cm 2cm 2cm]{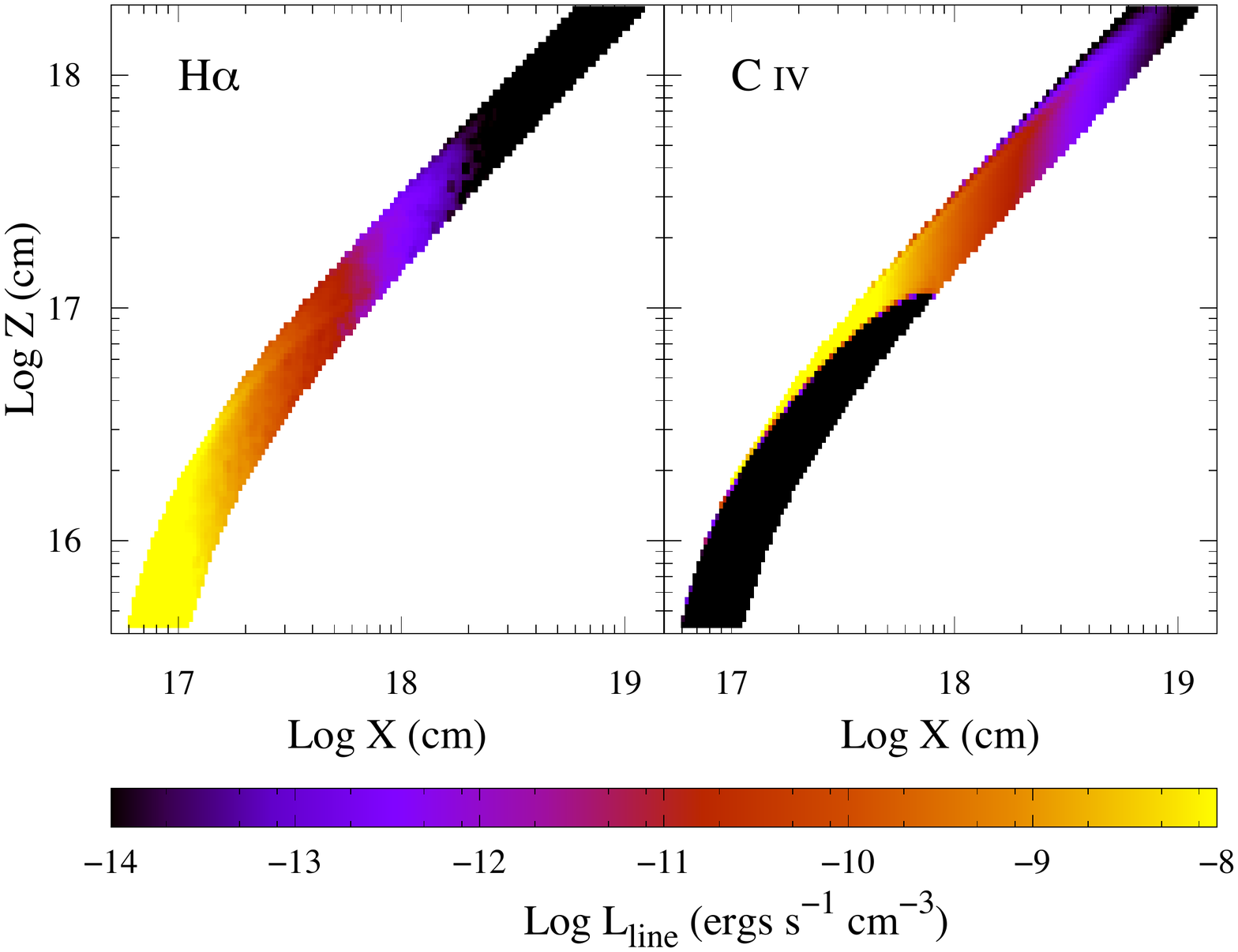}
      \caption{The line emissivity distribution for the H$\alpha$ and C~\textsc{iv} lines in the QSO model.}
      \label{fig:agn_lines}
    \end{minipage}
\end{figure*}

Figures~\ref{fig:tf_agn_ha} and~\ref{fig:tf_agn_c4} show
the 2-D EWRFs and response functions we
predict for the C~\textsc{iv} and H$\alpha$ lines in the $i=20^\circ$
QSO model. Figures~\ref{fig:agn_ions} and~\ref{fig:agn_lines} show the
distribution of the C~\textsc{iv} and H~\textsc{i} ionization fractions and
line luminosities within the outflow.

As shown in the upper panels of Figures~\ref{fig:tf_agn_ha} and~\ref{fig:tf_agn_c4}, the
EWRFs at short delays generally lie within the $v \propto \tau^{-1/2}$
envelope expected for a virialized flow dominated by Keplerian rotation, $v \propto R^{-1/2} \propto (c\tau)^{-1/2}$. The outflow kinematics of our models are only marginally apparent and only at the longest delays. In this regime the EWRFs, particular that for C~\textsc{iv}, show a weak diagonal "blue-leads-red" structure. However, the strongest overall feature is found at the shortest delays ($\tau \lesssim 50$~days) and positive velocities ($v \simeq 2000~{\rm km~s^{-1}}$). Partly because of this, the EWRF in the observationally relevant regime $\tau \lesssim 200$~days appears to have a diagonal slant towards short delays and positive velocities, which could easily be (mis-)interpreted as an {\em inflow} signature.


In addition, whilst the two EWRFs are broadly similar, the C~\textsc{iv} function is clearly narrower in velocity space. This reflects the difference in the respective line forming regions, as shown in Figure \ref{fig:agn_lines}: H$\alpha$ is concentrated in the Keplerian-dominated wind base, whilst C~\textsc{iv} has a substantial contribution from the extended outflow (as well as a contribution from a thin layer near the inner edge of the wind).

Broadly speaking, the {\em response} functions look similar to the EWRFs, but there are some important differences. For example, the response functions for the two lines differ more than their EWRF counterparts, with the long delay component being far stronger in C~\textsc{iv} than in $H\alpha$. Thus even though the C~\textsc{iv} {\em emission} is quite concentrated towards short delays, the {\em response} of the line is more evenly spread across a range of delays. Conversely, we see a weak \emph{negative} response at long delays in H$\alpha$. This happens because an increase in continuum luminosity over-ionizes the extended wind and pushes the line-forming region (seen in figure \ref{fig:agn_lines}) down towards the denser wind base. The potential to generate negative responses is an important feature of smooth/micro-clumped disc wind models. As we shall see in Section~\ref{sey}, it can have a very significant effect on the overall reverberation signature.

\subsection{A moderate luminosity Seyfert galaxy}
\label{sey}

Since the most extensive RM campaigns to date have been carried
out on nearby Seyfert galaxies, rather than on QSOs, it is important
to ask what the reverberation signature of a rotating disc wind
BLR would be in this setting. We can attempt to answer this question
by down-scaling down the QSO model developed by \citet{Matthews2016}.
In principle, there are many ways to carry out such a
scaling, since essentially all system and outflow parameters may be
mass- and luminosity-dependent. As a first step in this
direction, here we take the simplest possible approach and
scale the wind launching radii, wind acceleration length and
X-ray luminosity in the model in proportion with the adopted
BH mass ($10^7~M_\odot$), as suggested by \citet{Sim2010}.

In order to make our Seyfert model at least somewhat specific, we
also adopt $\dot{M}_{\mathrm{acc}}=0.02~M_\odot~\mathrm{yr}^{-1}$,
based on the inferred accretion rate of NGC~5548 \citep{Woo2002,Chiang2003}.
We then set $\dot{M}_w = \dot{M}_{\mathrm{acc}}$,
as in the QSO model, and the basic
geometry and kinematics of the outflow are also
kept exactly the same as for the QSO case.
The overall set of parameters describing our
Seyfert model are listed in Table~\ref{table:params}.
Even though this model
should not be construed as an attempt to {\em fit} observations of
NGC~5548, it is instructive as a point of comparison
for RM campaigns on this systems, such as that recently carried out by
the AGNStorm collaboration \citep{DeRosa2015a,Kriss2015,Edelson2015}.

\begin{figure}
	\includegraphics[height=\columnwidth, angle=-90, trim=1.5cm 4cm 1cm 1cm]{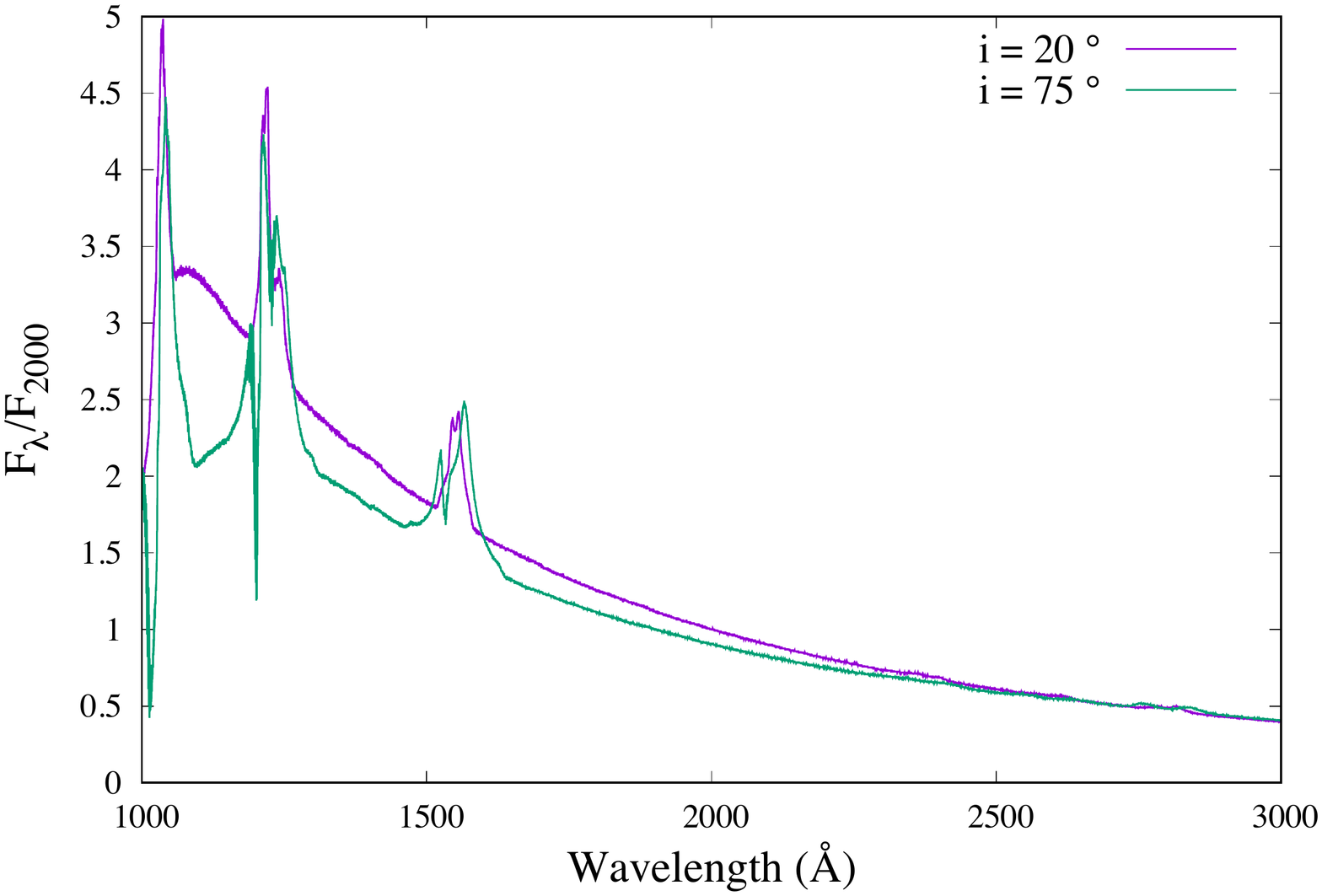}
    \caption{Spectra for Seyfert model along above wind $i = 20^{\circ}$ and through-wind (BAL-equivalent in QSO model) $i=75^{\circ}$ sightlines, for comparison to Figure \protect\ref{fig:spec_agn}, normalised to flux at 2000~\AA. Only very faint BAL features are apparent for BALQSO-like sightlines.}
    \label{fig:spec_sey}
\end{figure}

Figure~\ref{fig:spec_sey} shows the UV and optical spectra predicted by our
NGC~5548-like Seyfert I model for the same two sightlines as for our
QSO model. In this case, even the sightline looking directly into
the outflow does {\em not} produce a strong BAL signature in the UV resonance
lines. This is interesting: to the best of our knowledge, there are no
Type I Seyfert galaxies that exhibit permanent BALs, and only two that
have shown bona-fide transient BAL features. One of these is, in fact,
NGC~5548 \citep{Kaastra2014},
the other being WVPS~007 \citep{Leighly2009}.
Additionally, in both of those cases the BALs appeared
superposed on the overall BEL features without dipping significantly
below the local continuum level. Thus, empirically, BALs are
rare in Seyfert galaxies compared to QSOs. Intriguingly, our result
here suggests that the BLR in Seyferts may nevertheless be an outflow,
even one with the same overall geometry as those responsible for the
BALs in QSOs.

\begin{figure}
	\includegraphics[width=\columnwidth, trim=1cm 0cm 0cm 0cm]{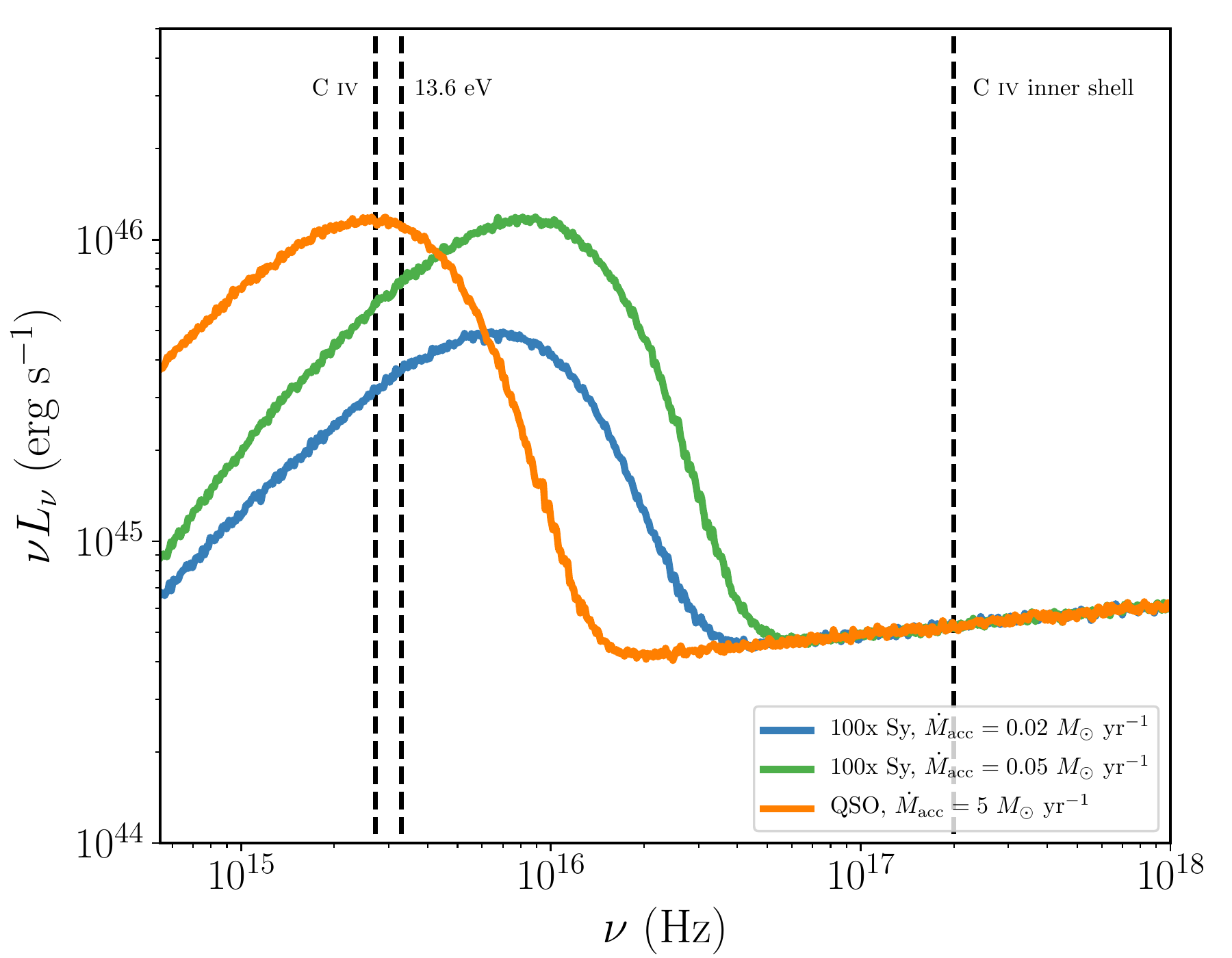}
    \caption{The spectral energy distributions (SEDs) for Seyfert and QSO models, including the contributions of both disc and corona. Our standard Seyfert model has an Eddington-normalized accretion rate of $\dot{m}_{acc} \simeq 0.1$, but we also show here an additional Seyfert model in which the Eddington fraction is identical to the QSO model, $\dot{m}_{acc} \simeq 0.2$. The SEDs for both Seyfert models are multiplied by a factor of 100, so they can be shown on the same scale as that for the QSO model. The locations of the \ion{H}{i} ($13.6$eV), \ion{C}{iv} and \ion{C}{iv} inner shell ionization edges are marked.}
  \label{fig:agn_sey_SED}
\end{figure}

Physically, the reason for the weakness of BAL features
in the Seyfert model is related to the ionization structure of the outflow. In this type of model, the more distant parts of the outflow that are responsible for producing BALs in the QSO model are overionized (relative to the ionization stages associated with BAL
transitions). The optical depth can therefore be insufficient for the production of strong, broad absorption features in these transitions,
even for sightlines that look directly into the outflow cone. This overionization arises because Seyferts preferentially harbour black holes at
the lower end of the mass range for AGN. As a result, their accretion discs extend to much smaller radii and reach much higher maximum temperatures. This leads to
the spectral energy distribution being noticeably harder in comparison to QSOs, as seen in Figure \ref{fig:agn_sey_SED}. The tendency of
AGN with lower mass BHs to produce more ionized winds has also been seen in hydrodynamic models of line-driven winds and reduces the efficiency of this
driving mechanism \citep{Proga2004, Proga2005}. This effect might explain the relative dearth of UV BAL features in Seyferts, as well as the occurrence of highly blue-shifted and highly ionized X-ray lines \citep{Pounds2014, Reeves2016, Tombesi2010}.

\begin{figure*}
	\begin{minipage}{.48\textwidth}
      \includegraphics[width=\columnwidth, trim=0cm 0cm 0cm 0cm]{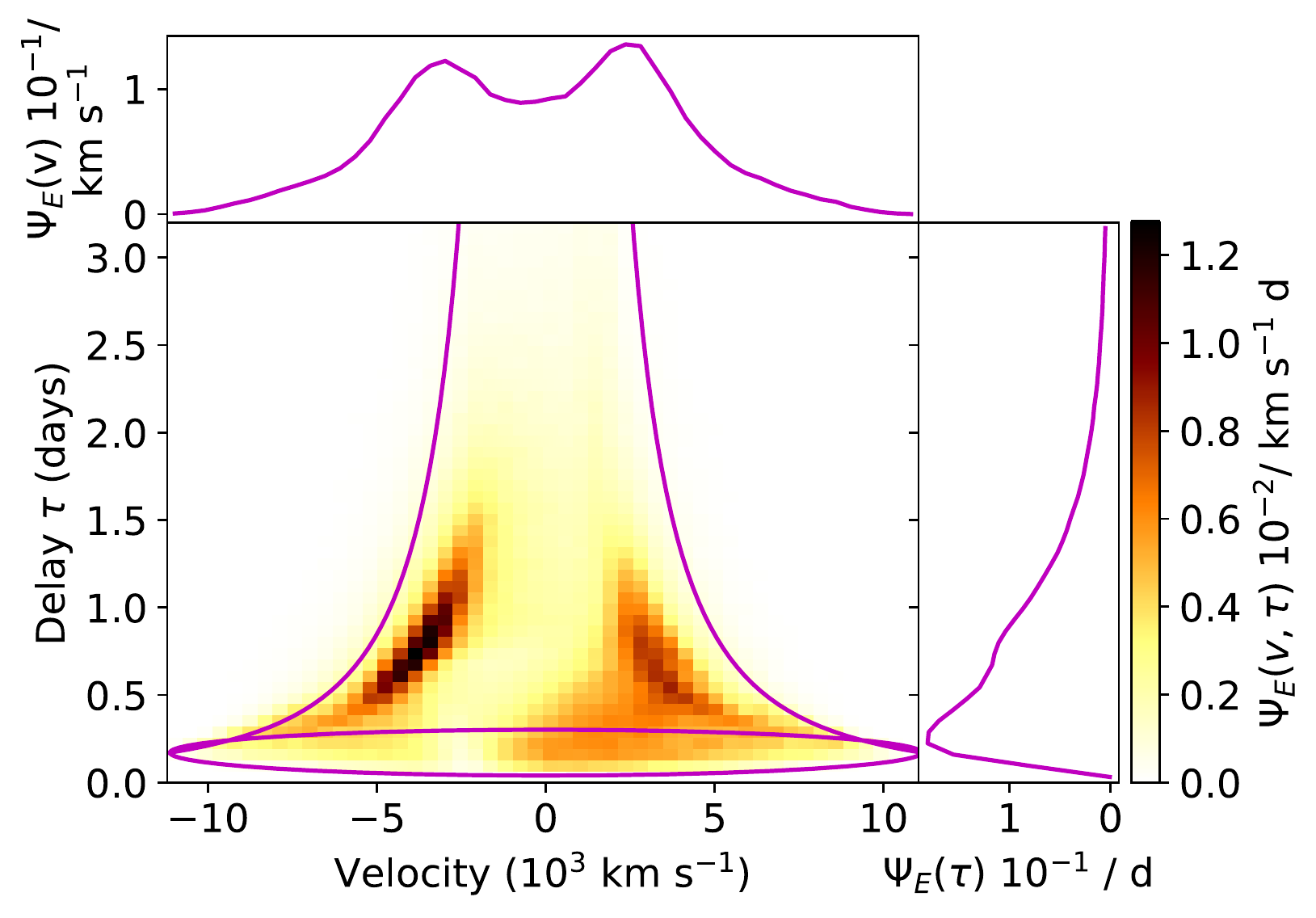}
      \includegraphics[width=\columnwidth, trim=0cm 0cm 0cm 0cm]{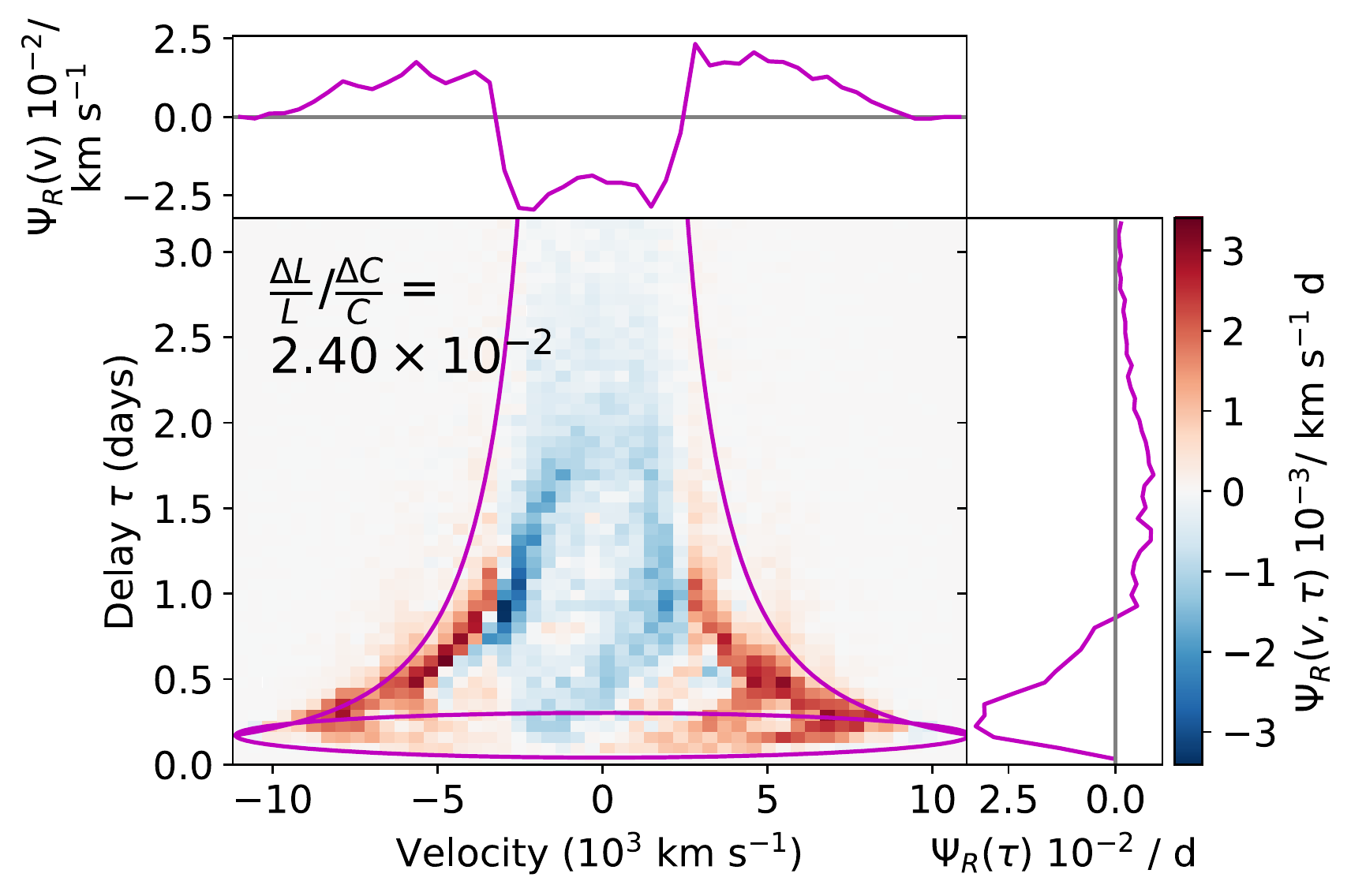}
      \caption{The velocity-resolved EWRF (upper) and response function (lower) for H$\alpha$ in
      the Seyfert model, shown for time delays up to 3.2 days. The violet lines illustrate the function edges for a Keplerian disc at the wind launching radius, and the $\nu \propto \tau^{-1/2}$ "virial flow" envelope.}
      \label{fig:tf_sey_ha}
    \end{minipage}\hfill
    \begin{minipage}{.48\textwidth}
      \includegraphics[width=\columnwidth, trim=0cm 0cm 0cm 0cm]{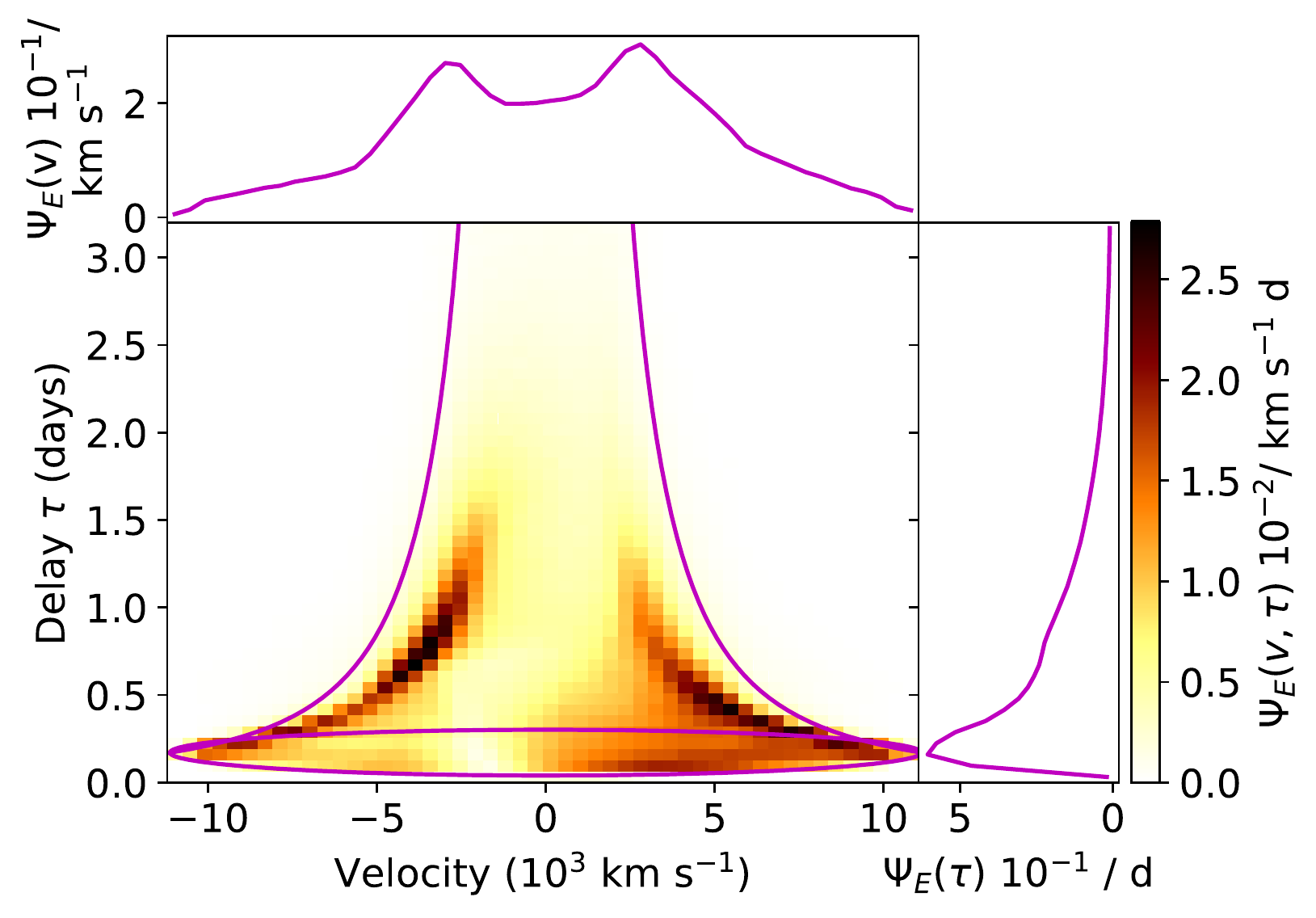}
      \includegraphics[width=\columnwidth, trim=0cm 0cm 0cm 0cm]{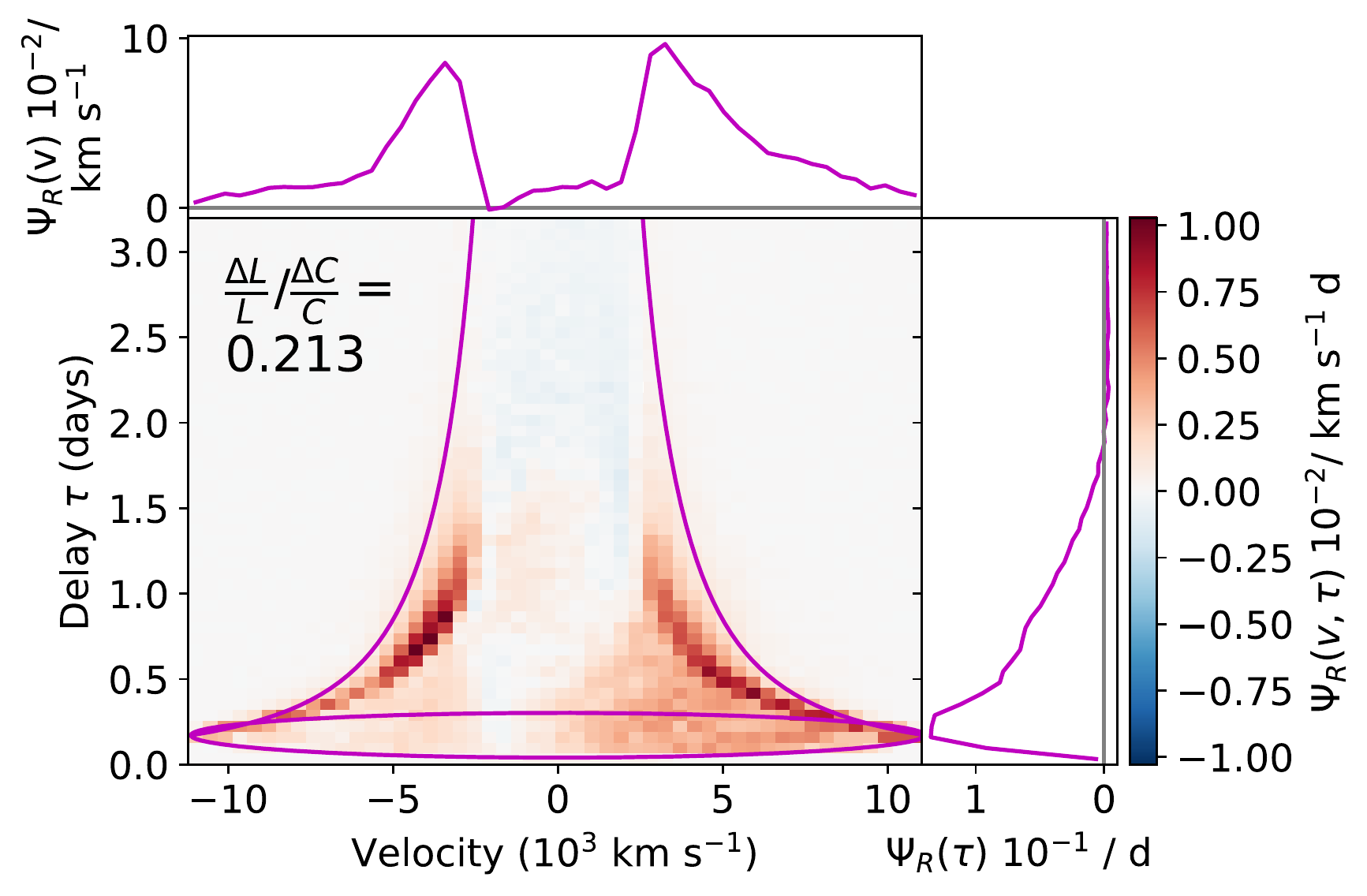}
      \caption{The velocity-resolved EWRF (upper) and response function (lower) for C~\textsc{iv} in
      the Seyfert model, shown for time delays up to 3.2 days. The violet lines illustrate the function edges for a Keplerian disc at the wind launching radius, and the $\nu \propto \tau^{-1/2}$ "virial flow" envelope.}
      \label{fig:tf_sey_c4}
    \end{minipage}

	\begin{minipage}{.48\textwidth}
      \includegraphics[width=\columnwidth, trim=2cm 4cm 2cm 2cm]{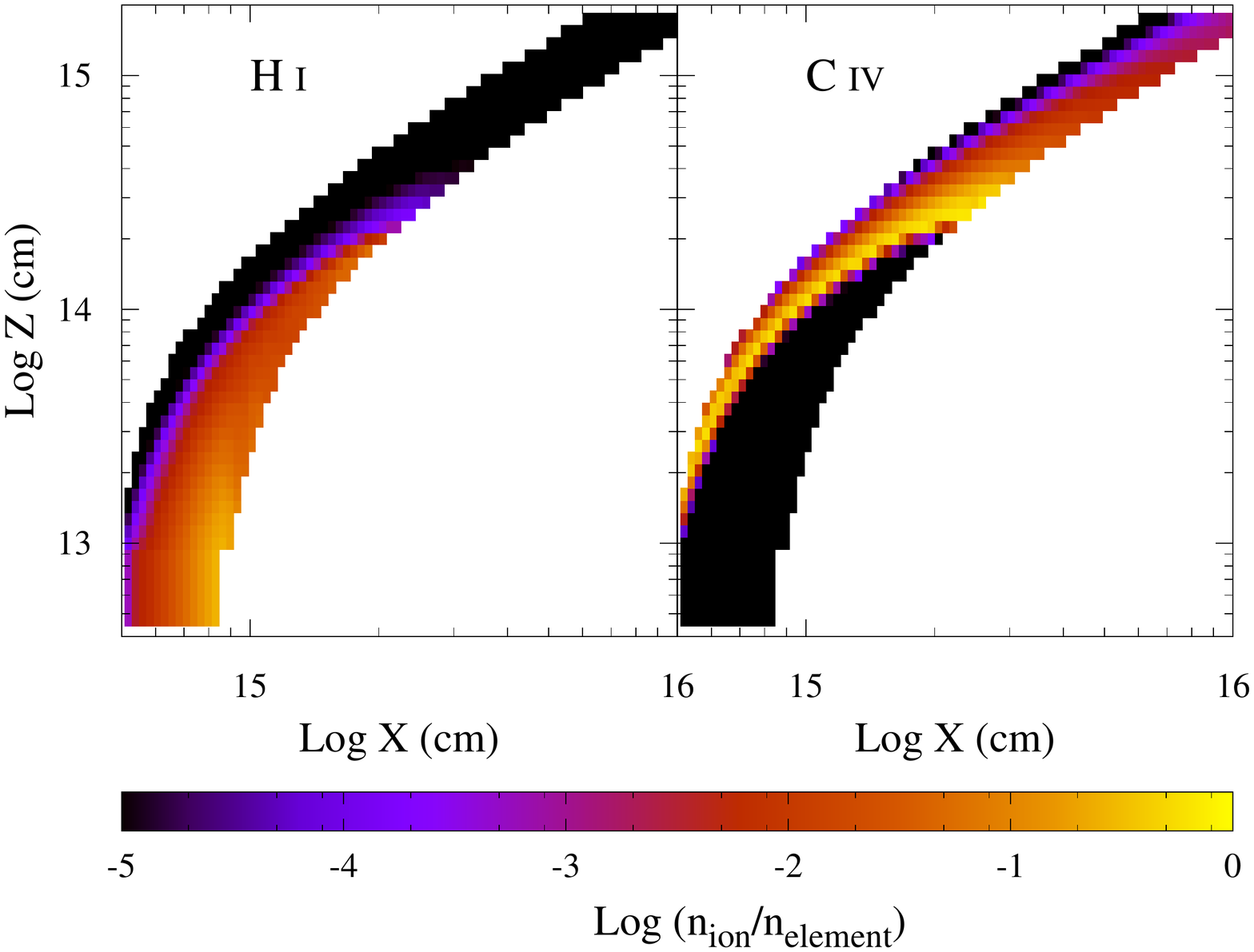}
      \caption{The H~\textsc{i} and C~\textsc{iv} ionization fractions within the wind in the Seyfert model. This graph shows a smaller region of the wind than figure \ref{fig:agn_ions} as the outflow is overionized in the Seyfert model and does not show emission in common RM lines.}
      \label{fig:sey_ions}
    \end{minipage}\hfill
    \begin{minipage}{.48\textwidth}
      \includegraphics[width=\columnwidth, trim=2cm 4cm 2cm 2cm]{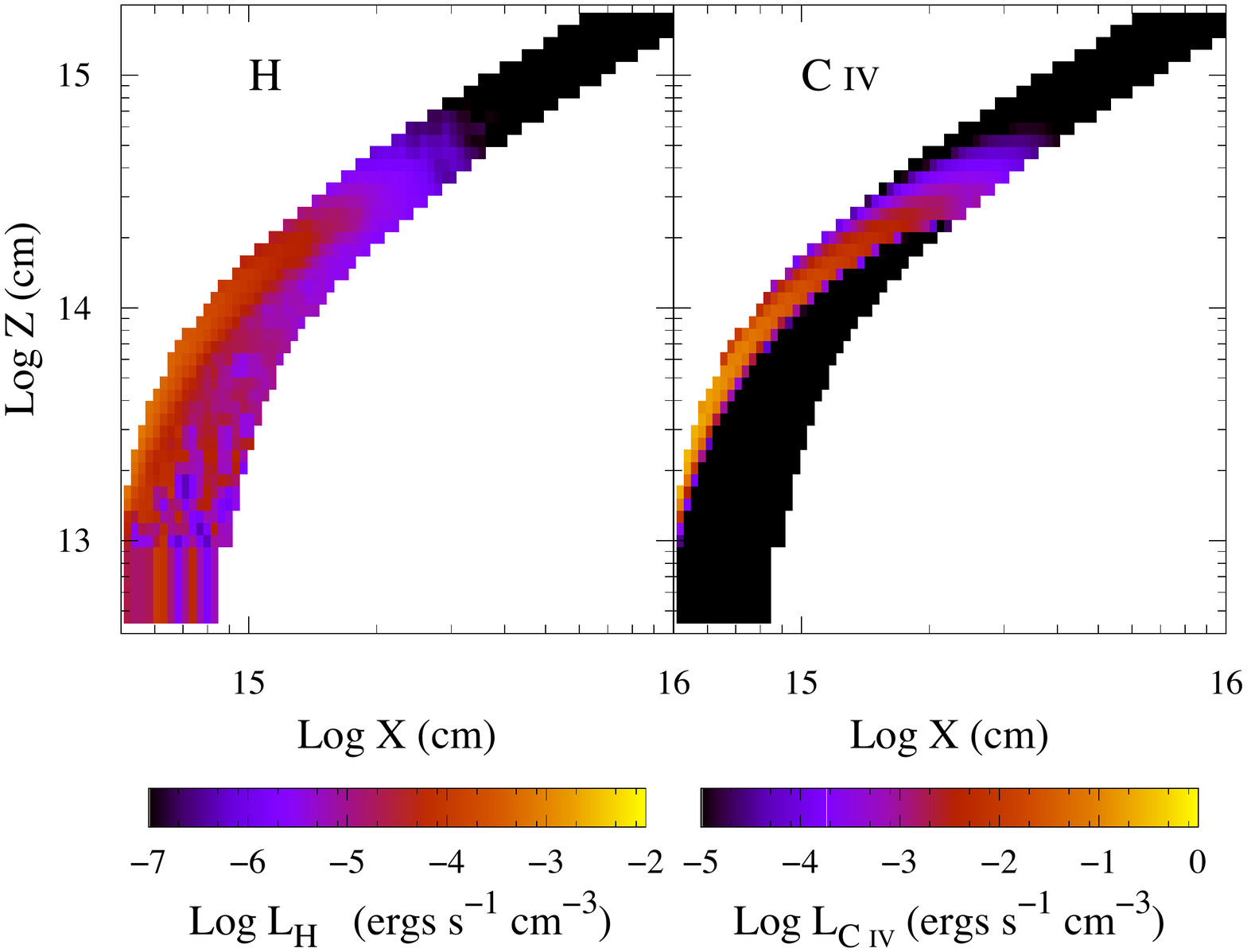}
      \caption{The line emissivity distribution for the H$\alpha$ and C~\textsc{iv} lines in the Seyfert model. This graph shows a smaller region of the wind than figure \ref{fig:agn_lines} as the outflow is overionized in the Seyfert model and does not show emission in common RM lines.}
      \label{fig:sey_lines}
    \end{minipage}
\end{figure*}

Figures~\ref{fig:tf_sey_ha} and~\ref{fig:tf_sey_c4} show the predicted 2-D EWRFs and response
functions for H$\alpha$ and C~\textsc{iv} for our
NGC~5548-like model Seyfert I model.
To aid interpretation, we also once again show the distribution of the
H~\textsc{i} and C~\textsc{iv} ionization fractions and
line luminosities within the outflow (Figures~\ref{fig:sey_ions}
and~\ref{fig:sey_lines}).

The EWRFs in Figures \ref{fig:tf_sey_ha} and \ref{fig:tf_sey_c4} show reverberation signatures roughly similar to those of QSO models. However, the characteristic delay timescales are much shorter, as would be expected from the much smaller wind launch radius ($1/100$\textsuperscript{th} that of the QSO model). This trend is also observed empirically (\citealt{Bentz2013}; also see Section~\ref{sec:lum_t} below). There are also other differences from the QSO EWRFs. Perhaps most notably, the velocity ranges in which C~\textsc{iv} and H$\alpha$ are preferentially found have switched: H$\alpha$ now appears at lower characteristic velocities than C~\textsc{iv}. The reason for this is clear from Figure \ref{fig:sey_ions}. The inner edge of the disc wind in these models is more strongly ionized in the Seyfert model compared to the QSOs one. This pushes the H$\alpha$ emission back into the dense, heavily-shielded back of the wind. Meanwhile, C~\textsc{iv} emission is concentrated in a thin ionization front near the inner edge of the flow and along the transition region between the dense, quasi-Keplerian base and the fast wind. The outflow-dominated distant regions of the flow are strongly over-ionized, so there are no obvious diagonal outflow signatures in Figures \ref{fig:tf_agn_ha} and \ref{fig:tf_agn_c4}, even at long delays.

The {\em response} functions for the Seyfert model are even more interesting. Most importantly, both H$\alpha$ and C~\textsc{iv} exhibit strong {\em negative} responses in the low-velocity regime ($|v| \lesssim 5000~{\rm km~s^{-1}}$). In the case of H$\alpha$, this negative response contribution is strong enough to render even the integrated line response negative. Unlike in LOC-type models, negative responsivities can arise naturally in smooth/micro-clumped models. A simplistic explanation for this behaviour is that, as the continuum luminosity increases or decreases, any given line-forming region tend to move outwards or inwards. Parts of the wind that shift {\em out of} the line-forming region when the continuum increases will exhibit negative responsivities. In practice, the line-forming region is, of course, not always so sharply delineated.

\begin{figure*}
      \includegraphics[width=.98\columnwidth, trim=0.8cm 0cm 0.8cm 0cm]{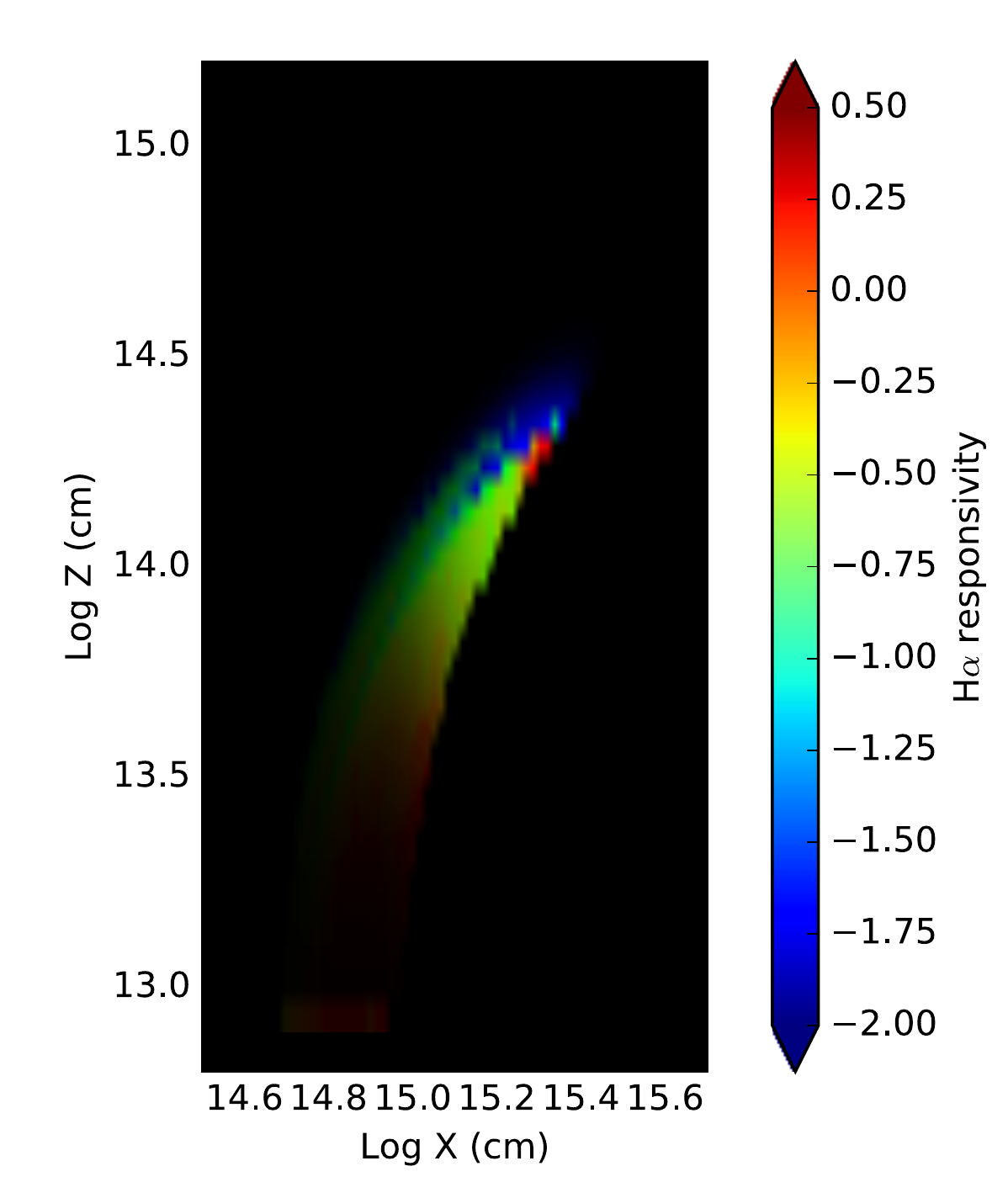}\hfill
      \includegraphics[width=.98\columnwidth, trim=0.8cm 0cm 0.8cm 0cm]{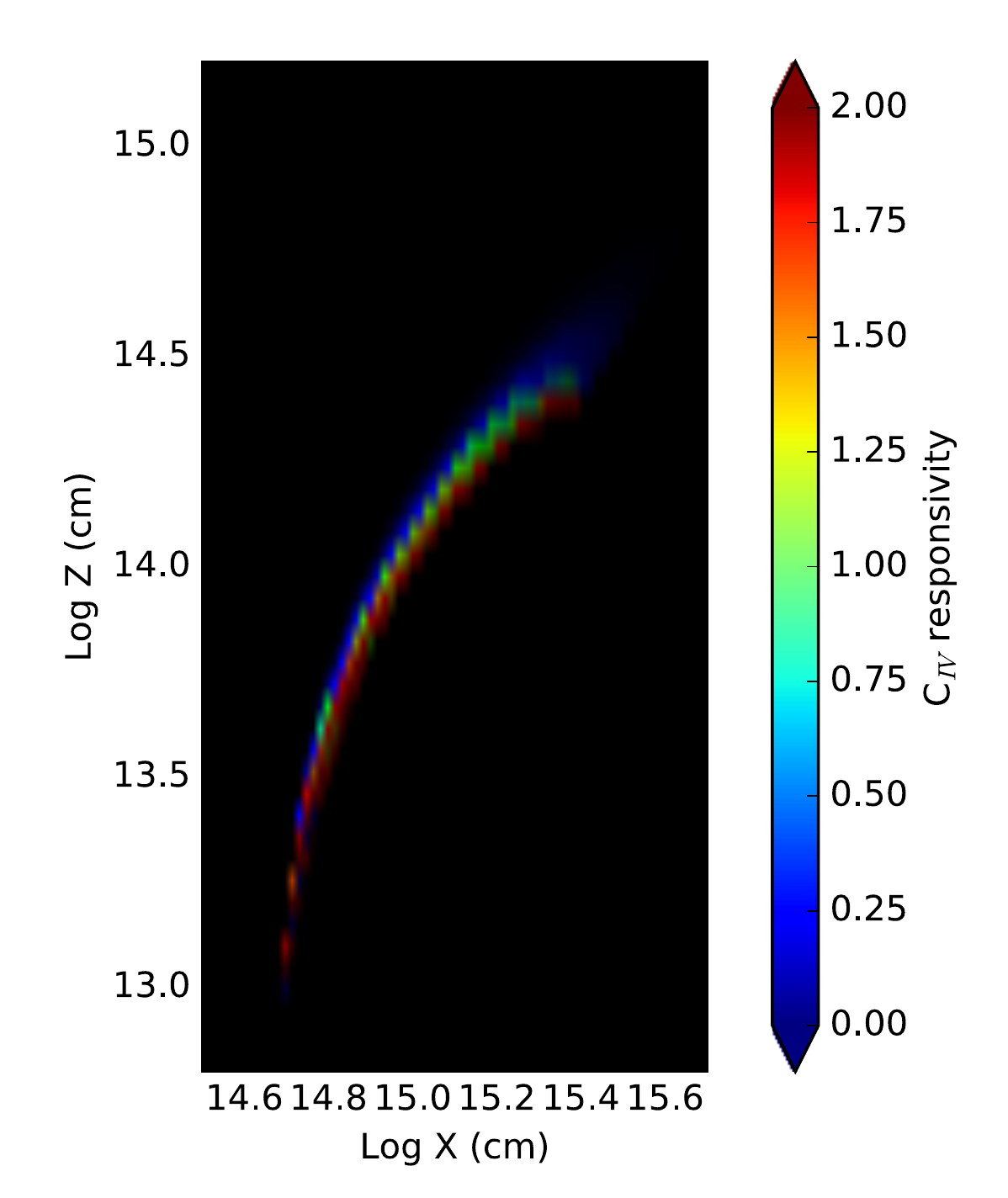}
      \caption{The local responsivity $\eta$ for H$\alpha$ (left) and C~\textsc{iv} (right). Colour indicates responsivity, brightness indicates local emissivity. The brightest cells on the figure are in the upper 99.9\% of cells for line emission; other cell brightnesses are scaled linearly to this value.}
      \label{fig:sey_eta}
\end{figure*}

To illustrate this, Figure \ref{fig:sey_eta} shows the local responsivities in the wind. The C~\textsc{iv} line provides a  particularly simple example, since here the emission is concentrated near a thin ionization front that shifts backwards through the wind with increasing luminosity. In this case, the regions of peak emission all exhibit positive responsivities of order $\eta \simeq 1$. By contrast, there is no clear ionization front for H$\alpha$, which is instead mostly produced near the outer edge of the wind. Near the disc plane, in the dense base of the wind, the responsivity remains positive, corresponding to the positive response seen at high velocities in Figure \ref{fig:tf_sey_ha}. However, the responsivity is mostly negative everywhere else. In these regions, the increased continuum luminosity has changed the thermal and ionization state in such a way as to {\em reduce} the number of photons that are produced.

Seyfert galaxies have long been attractive targets for RM campaigns, thanks to their short characteristic delays. Our results suggest that such campaigns may be able to test models such as ours, but considerable care needs to be taken in interpreting the observational data. Most importantly, attempts to reconstruct the response function from such data sets need to allow for the possibility of {\em negative} emission line responses in parts of the velocity-delay plane. Moreover, it is {\em not} safe to assume that all rotating disc winds will produce response functions with a diagonal structure running from negative velocities and short delays to positive velocities and long delays. Such a structure is commonly thought of as the smoking-gun of an outflow \citep{Denney2009}, but our disc wind models do not produce this.



\section{Discussion}
\label{sec:Discuss}

In the previous section, we have presented the reverberation signatures predicted by smooth/micro-clumped rotating disc winds for representive QSO and Seyfert galaxy models. The resulting response functions are complex and differ substantially from those predicted by simple models for rotating accretion discs and quasi-spherical outflows \citep{Welsh1991}.

Perhaps most importantly, at the short lags that dominate the overall response, our response functions do {\em not} usually show the ``blue-leads-red" diagonal structure
that is often considered to be a characteristic outflow signature (cf. Figure~\ref{fig:tf_example}). Such a signature is really only visible in our QSO model for C~\textsc{iv}, on timescales of $\gtrsim 100$ days. Given the limitations of current reverberation mapping campaigns, this is unlikely to be detectable in practice. Instead, the response functions for our Seyfert models are roughly symmetric, indicating that rotational kinematics dominate in the line-forming regions. This finding is in line with earlier disc-wind modelling efforts \cite{Chiang1996, Kashi2013, Waters2016}.

In our QSO models, the strongest response actually occurs at short lags and positive velocities. Indeed, the overall response functions for this model could easily be (mis-)interpreted as the "red-leads-blue" signature of an {\em inflow} (see Figures~5 and 6). "Red-leads-blue" reverberation signatures have, in fact, already been observed in several AGN and interpreted as evidence for inflows \citep{Grier2012, Bentz2010,Gaskell1988,Koratkar1989}. Based on our results, we urge caution in taking such interpretations at face value, at least in the absence of detailed kinematic and radiative transfer modelling of the data.

Another key feature of our disc-wind reverberation modelling is the prevalence for {\em negative} responses, particularly at low velocities and long delays. Negative responsivities are a common occurrence in smooth/micro-clumped flows, as an increase in the ionizing continuum shifts the line-forming region through the flow, enhancing the emissivity in some regions, but reducing it in others. This is similar to the way in which negative responsivities can arise in "pressure-law" models containing a significant portion of optically thin clouds (Goad, O'Brien \& Gondhalekar 1993). As such, we predict that negative responses will arise in a wide range of systems, not just our specific model.

The process of inferring response functions from observational data is sensitive to negative responsivities in parts of the velocity-delay plane. At least one technique that allows for this possibility is available (Regularized Linear Inversion: \citet{Krolik1995a, Skielboe2015}). However, both of the two most widely used methods -- maximum entropy inversion \citep{Krolik1991, Horne1991, Horne1994, Ulrich1996, Bentz2010, Grier2013} and forward dynamical modelling \citep{Pancoast2011, Waters2016} -- explicitly assume that the responsivity is positive-definite. Given that negative responsivities are quite common in smooth/micro-clumped BLR models, this may not be a safe assumption.

We have also pointed out the weakness of BAL features in the spectra we have calculated for our Seyfert model, even for sightlines that look directly into the wind cone. This same sightline does produce strong BAL features in our QSO model. This behaviour is consistent with the observational dearth of BAL features in Seyfert galaxies. In our models, the absence of strong BALs in Seyferts is associated with the higher ionization state of their outflows. This is interesting, since it implies that disc winds might be ubiquitous in Seyfert galaxies -- and even dominate the BLR emission -- even though hardly any of them show the BAL features that are the smoking gun for such outflows in QSOs.

\subsection{Comparison to empirical lag-luminosity relations}
\label{sec:lum_t}

A detailed comparison of our predicted response functions to
observations is beyond the scope of the present study. However, it is
instructive to ask whether the characteristic delays we predict for
the C~\textsc{iv} and H$\beta$ lines are broadly consistent with the
trends that have been established empirically \citep{Peterson2005,
Peterson2006a, Bentz2013}.

An important consideration here is how to calculate the appropriate characteristic delay. In observational analyses, lags are usually estimated from the cross-correlation function (CCF) of the line and continuum light curves. More specifically, a typical procedure is to quote the centroid of the CCF, calculated within some fixed distance from the peak of the CCF. As discussed in detail by Welsh (1999), the CCF is equal to the velocity-integrated 1-D EWRF/response function convolved with the auto-correlation function of the continuum. Thus the CCF is simply a blurred version of the 1-D EWRF/response function. Consequently, we estimate characteristic delays for our models by calculating the centroid of the EWRF/response function within 80\% of its peak (cf \citet{DeRosa2015}).
\footnote{It is worth noting that centroid delays calculated across the {\em entire} EWRF can be significantly longer, by as much as $\times 3$ or more. Moreover, the overall response of the H$\alpha$ line in our Seyfert model -- integrated over both velocity and delay -- is negative, so it is hard to justify any estimate of a  a global centroid delay for this line.}

\begin{figure*}
    \includegraphics[width=\textwidth, trim=0cm 0cm 0cm 0cm]{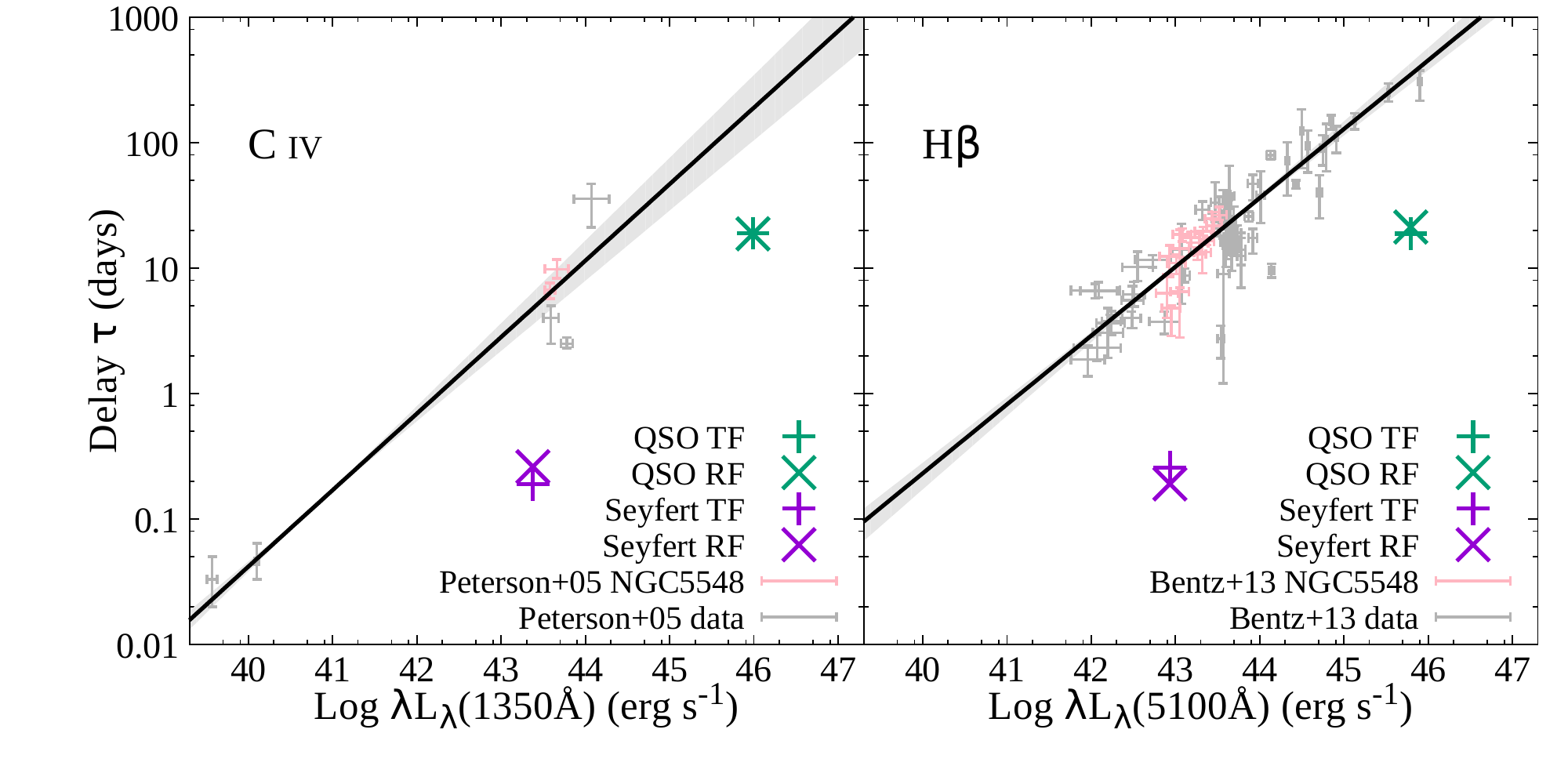}
    \caption{Luminosity-delay relationship for $H\beta$ and C~{\sc iv} compared to that determined observationally by \protect\citet{Bentz2013} and \protect\citet{Peterson2004}. The line and shaded area around it are the fit and error on the fit. The points are the values from \protect\citet{Bentz2013} and \protect\citet{Peterson2004} with errors, with those corresponding to NGC5548 highlighted.}
    \label{fig:lum_t}
\end{figure*}

Figure~\ref{fig:lum_t} shows the empirical lag-luminosity relations for both transitions, over a luminosity range that includes both Seyferts and QSOs. The luminosities and mean lags predicted by our models are also shown. The luminosities are those that would be inferred from the spectrum observed for this particularly sightline (i.e. $\lambda L_{\lambda} = 4 \pi d^2 \lambda F_{\lambda}$). For comparison, we show delays estimated from both EWRFs and response functions for our standard Seyfert and QSO models.

Figure~\ref{fig:lum_t} shows that the lags predicted by our standard Seyfert and QSO models are too small to match observations, by about an order of magnitude. The {\em slope} of the line connecting low- and high-luminosity models matches the data very well, however. Thus our models reproduce the observed power-law index of the lag-luminosity relation, but not its normalization. In principle, the simplest way to reconcile models and observations would be to move the wind-launching region radially outwards in the accretion disc. In practice, this will also require other  parameter changes in order to roughly preserve the ionization state and spectral signatures of the outflow.

\subsection{The inclination dependence of the virial product}
\label{sec:disc:angle}

As mentioned in Section~\ref{sec:background}, RM-based SMBH mass estimates rely
on the determination of the so-called {\em virial product}, $c \tau_{BLR} v_{BLR}^2  / G$.
Even if the velocity field in the BLR is
dominated by the gravitational field of the SMBH, the virial product
is, strictly speaking, not an estimate of $M_{BH}$, but is only
proportional to it,
\begin{equation}
	M_{BH} = f c \tau_{BLR} v_{BLR}^2  / G,
	\label{eqn:virial}
\end{equation}
where $f$ is the constant of proportionality (the so-called "virial factor").
The need to introduce
$f$ arises primarily because the BLR is unlikely to be spherically
symmetric. As a result, the measured lags and velocities are both
expected to depend on the observer orientation with respect to the
system. In such situations, $f$ will then also depend on the emission
line that is being used, due to the ionization stratification of the
BLR.

\begin{figure*}
	\includegraphics[width=0.8\textwidth, trim=0cm 0cm 0cm 0cm, angle=0]{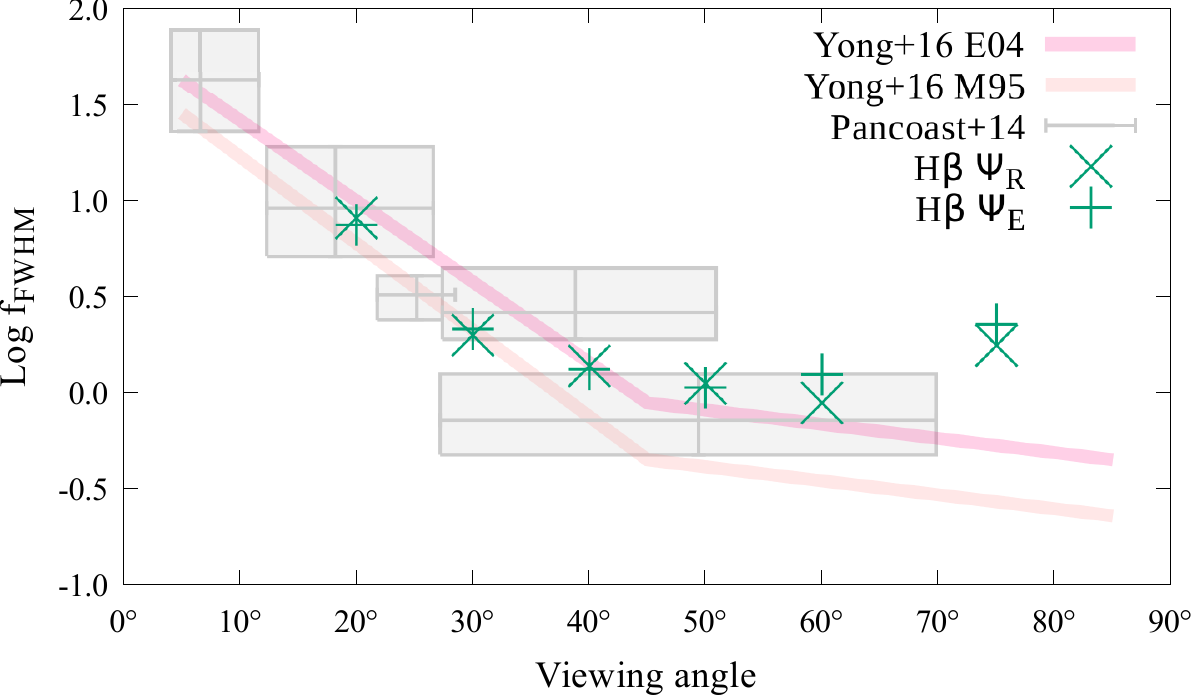}
    \caption{The virial factor (Equation \protect\ref{eqn:virial}) for our QSO disc wind model. The values listed here have been estimated from the velocity widths (FWHM) and centroid delays of the emission lines for mean emission. Observational data for 4 AGN from \protect\citet{Pancoast2014} are plotted with their error bars. Two sets of models from \protect\cite{Yong2016} are also shown (see text for details).}
    \label{fig:f_by_angle}
\end{figure*}

There have been several attempts to estimate a ``typical'' or mean
value for $f$. The derived values typically lie in the range $3
\lesssim f \lesssim 6$ \citep[e.g.][]{Onken2004, Woo2010, Park2012, Grier2013, Ho2014}. More recently, \citet{Pancoast2014} carried out an inverse RM analysis for several
AGN in order to constraint the geometry and kinematics of the BLR. The
inclination-dependence of the virial product -- i.e. $f(i)$ -- is a
byproduct of their analysis and is shown in Figure~\ref{fig:f_by_angle}.
We also show there the values of $f$ predicted by \citet{Yong2016}
from two types of disc wind models in the literature \citep{Murray1995, Elvis2004},
for two specific inclinations. All of these results are for the H$\beta$ line.

For our own models, $M_{BH}$ is an input parameter, and we can measure
$v_{BLR}$ from our predicted spectra and $\tau_{BLR}$ from our
reverberation modelling. We can therefore easily calculate $f(i)$ for
the models. The results for our $M_{BH} = 10^9~M_{\odot}$ QSO model are given in Table \ref{table:angles}, where we provide our estimates of the virial factor associated with the H$\beta$ line for five viewing angles. In Figure~\ref{fig:f_by_angle}, we compare these calculations to the results of \citet{Pancoast2014} and \citet{Yong2016}. \citet{Pancoast2014} model observations of AGN with  $10^6 \lesssim M_{BH} \lesssim 10^7 M_\odot$, whilst \citet{Yong2016} adopt $M_{BH}=10^8 M_\odot$ for their models.
Despite these differences in $M_{BH}$, we find reasonable agreement between
all of these estimates.

We do see one feature in our calculations of the virial factor that is not present in the simulations of \citet{Yong2016}. At angles approaching the edge of the wind cone, $f$ increases markedly in our models. This increases arises as sightlines close to the disc wind see their blue wing suppressed significantly, narrowing the line and reducing the estimated $v_{BLR}$. Sightlines through the disc wind experience an even greater suppression of their blue wing, but this is counterbalanced by an increase in the associated delay. Real AGN disc winds are unlikely to have sharp edges, of course. In any case, there is currently no observational data at high inclinations that could be used to look for such a feature.

\begin{table}
	\caption{Table of f factors against observation angles.}
	\label{table:angles}
	\begin{tabular}{r | r r r r}
    	\hline \hline
						& \multicolumn{2}{c}{Log f}	\\
		Angle			& H$\beta$ $\Psi_{R}$ 	& H$\beta$ $\Psi_{E}$  \\
							\hline
		$20^{\circ}$	& 0.9102				& 0.8738				\\
		$30^{\circ}$	& 0.3011				& 0.3330				\\
		$40^{\circ}$	& 0.1410				& 0.1239				\\
		$50^{\circ}$	& 0.0509				& 0.0293				\\
		$60^{\circ}$	& -0.0508				& 0.0965				\\
		$75^{\circ}$	& 0.2478				& 0.3573				\\
        \hline \hline
	\end{tabular}
\end{table}

\section{Summary}
\label{sec:Summary}
Reverberation mapping has become an important tool for studying the structure of the broad line region of AGN.  Here,  we have used a Monte Carlo radiative transfer code to calculate the reverberation signatures for rotating disc wind models of the BLR. The QSO model presented by \citet{Matthews2016} is used to describe high-luminosity AGN, while a scaled-down version of this model (with parameters similar to NGC~5548) is used to describe moderate-luminosity Seyferts. We present spectra, velocity-delay maps and lag-luminosity relationships for the H$\alpha$ and C~{\sc iv} emission lines for these models. Our main conclusions are as follows:
\begin{itemize}
\item Smooth/micro-clumped disc-wind models of the BLR can produce {\em negative} responses in parts of the velocity-delay plane. This possibility needs to be taken into account when reconstructing response functions from observational data.
\item The kinematics of the line-forming region tend to be rotation-, rather than outflow-dominated. In some lines, and at short delays, the red wing can even lead the blue wing, a signature usually thought of as characteristic of {\em inflow}.
\item The classic "blue-leads-red" outflow signature can usually only be observed in the long-delay limit.
\item The slope of the lag-luminosity relations predicted by the models is consistent with observations, but the centroid delays are too short, by about an order of magnitude.
\item The dependence of the "virial product" on viewing angle in our models is consistent with the relationship suggested by recent observational modelling efforts.
\end{itemize}
Whilst our conclusions have been derived for a specific disc wind model, many of them  can be applied more generally.
\section*{Acknowledgements}
We are grateful to the anonymous referee for an excellent and constructive report that led to significant improvements to the paper.
We acknowledge the University of Southampton's Institute for Complex
Systems Simulation and the EPSRC for funding this research. CK
acknowledges support from the Leverhulme Foundation in the form of a
Research Fellowship and from STFC via a Consolidated Grant to the
Astronomy Group at the University of Southampton. Part of his
contribution to the project was carried out at the Aspen Center for
Physics, which is supported by National Science Foundation grant
PHYS-1066293. JHM is supported by STFC grant ST/N000919/1.
KSL acknowledges the support of NASA for this work through grant
NNG15PP48P to serve as a science adviser to the Astro-H project.




\bibliographystyle{mnras}
\bibliography{mybib.bib}


\bsp	
\label{lastpage}
\end{document}